\documentclass[11pt]{article}
\pdfoutput=1

\usepackage{cite}
\usepackage{booktabs}
\usepackage[english]{babel}
\usepackage{amsmath,amssymb,amsbsy,amstext, amsthm, simplewick}
\usepackage{hyperref}
\usepackage{graphicx}
\usepackage{amsfonts}
\usepackage{amssymb}
\usepackage[small]{caption}
\usepackage{upgreek}
\usepackage[svgnames,dvipsnames,x11names,table]{xcolor}
\usepackage{multirow} 
\usepackage{geometry}
\usepackage[hang,flushmargin]{footmisc}
\usepackage{bm}
\usepackage{braket}
\usepackage{subcaption}
\usepackage{simpler-wick}
\usepackage{mathtools}

\usepackage{colortbl}

\makeatletter
\newlength{\apb@width}
\newcommand{\autoparbox}[2][c]{\settowidth{\apb@width}{#2}\parbox[#1]{\apb@width}{#2}}

\makeatother

\definecolor{lightgray}{gray}{0.9}

\usepackage[framemethod=default]{mdframed}
\newmdenv[skipabove=7pt,
skipbelow=7pt,
rightline=false,
leftline=false,
topline=false,
bottomline=false,
backgroundcolor=gray!10,
linecolor=gray,
innerleftmargin=5pt,
innerrightmargin=5pt,
innertopmargin=5pt,
innerbottommargin=5pt,
leftmargin=0cm,
rightmargin=0cm,
linewidth=4pt]{eBox}

\numberwithin{equation}{section}

\def\beq{\begin{equation}}
\def\eeq{\end{equation}}

\def\bea{\begin{eqnarray}}
\def\eea{\end{eqnarray}}

\def\beq{\begin{equation}}
\def\eeq{\end{equation}}
\def\bea{\begin{eqnarray}}
\def\eea{\end{eqnarray}}

\def\O{{\cal O}}

\def\I{{\cal I}}
\def \K {{k}}
\def \bK {{\bm K}}

\def\k{{\vec k}}
\def\bk{{\bm k}}

\def\p{{\vec p}}

\def\x{{\vec x}}
\def\bx{{\bm x}}
\def\X{{x}}
\def\y{{\vec y}}
\def\by{{\bm y}}
\def\P{{p}}

\DeclareRobustCommand{\SkipTocEntry}[4]{}

\definecolor{blue3}{RGB}{31, 119, 180}
\definecolor{red3}{RGB}{	214, 39, 40}
\definecolor{orange3}{RGB}{255, 127, 14}
\definecolor{green3}{RGB}{44, 160, 44}

\begin{document}

\begin{titlepage}
\setcounter{page}{1} \baselineskip=15.5pt 
\thispagestyle{empty}

\begin{center}
{\fontsize{18}{18} \bf Dynamical RG and Critical Phenomena \\ \vskip 5pt in de Sitter Space}
\end{center}

\vskip 20pt
\begin{center}
\noindent
{\fontsize{12}{18}\selectfont  Daniel Green and Akhil Premkumar}
\end{center}

\begin{center}
\vskip 4pt
\textit{ Department of Physics, University of California, San Diego, La Jolla, CA 92093, USA}

\end{center}

\vspace{0.4cm}
 \begin{center}{\bf Abstract}
 \end{center}
 
 \noindent

Perturbative quantum field theory in de Sitter space is known to give rise to a variety of contributions that diverge with time (secular terms).   Despite significant progress, a complete understanding of the physical origin of these divergences remains an outstanding problem.  In this paper, we will study the origin of secular divergences in de Sitter space for interacting theories that are near attractive conformal fixed points.  We show that the secular divergences are determined by the anomalous dimensions of the same theory in flat space and can be re-summed using the dynamical renormalization group. This behavior is mandatory at the conformal fixed point but we show that it holds away from the fixed point as well.  We analyze this problem in general using conformal perturbation theory and study conformally coupled scalar fields in four and $4-\epsilon$ dimensions as examples.  

\end{titlepage}

\restoregeometry

\newpage
\setcounter{tocdepth}{2}
\tableofcontents

\newpage

\section{Introduction}

The physics of de Sitter space has posed both conceptual and technical challenges to our understanding of the universe.  Ultimately, the lack of a fixed boundary in the presence of dynamical gravity is an unavoidable challenge in defining physics in de Sitter space~\cite{Bousso:1999cb,Strominger:2001pn,Witten:2001kn,Mazur:2001aa,Maldacena:2002vr,Alishahiha:2004md}.  Yet, there have long been more mundane challenges associated with divergences in perturbation theory~\cite{Ford:1984hs,Antoniadis:1985pj,Starobinsky:1986fx,Tsamis:1994ca,Tsamis:1996qm,Tsamis:1997za,Weinberg:2005vy,Weinberg:2006ac,Senatore:2009cf,Polyakov:2012uc},  with and without dynamical gravity. Resolving the origin of these divergences is a more tangible problem than quantum gravity in de Sitter itself and is surely  a necessary step towards a complete theory of (quantum) cosmology.

Significant progress has been made in our understanding of perturbation theory in cosmological spacetimes.  In the particular case of cosmological correlators in single-field inflation, it has been shown to all-loop order that time-dependent contributions from individual diagrams must vanish when all the diagrams are summed together~\cite{Senatore:2012ya,Assassi:2012et}.  In essence, in the case of single-field inflation, the secular divergences are not physical. The absence of such divergences is unique to the metric fluctuations due to the nonlinear symmetries of these modes~\cite{Weinberg:2003sw,Hinterbichler:2012nm}, which are also responsible for the separate universe description of cosmology~\cite{Salopek:1990jq} and the single-field consistency conditions~\cite{Maldacena:2002vr,Creminelli:2004yq}.

In contrast to the metric fluctuations, the divergences associated with conventional quantum fields in cosmological backgrounds are physical~\cite{Weinberg:2006ac,Burgess:2010dd,Marolf:2010zp,Marolf:2011sh,Marolf:2012kh,Anninos:2014lwa, Burgess:2015ajz,Gorbenko:2019rza,Baumgart:2019clc}.  In some cases, these divergences can be re-summed using known techniques from quantum field theory (QFT)~\cite{Starobinsky:1986fx,Tsamis:2005hd,Riotto:2008mv,Seery:2009hs,Serreau:2013psa,Nacir:2016fzi}.  The dynamical renormalization group (DRG)~\cite{Tanaka:1975ti,Boyanovsky:1998aa,Boyanovsky:2003ui} is one such approach that is well-suited for secular divergences in cosmology~\cite{Boyanovsky:2004gq,McDonald:2006hf,Podolsky:2008qq,Burgess:2009bs}.  These divergences are at most logarithmic in the scale factor~\cite{Weinberg:2005vy,Weinberg:2006ac} and thus the leading logs can be resummed using the DRG. Unfortunately, the meaning of this resummation lacks the clear physical interpretation that we associate to the renormalization group in flat space.   

In recent work, a number of non-trivial features of tree-level perturbation theory in de Sitter space have been connected to conventional physics in flat space. For instance, the analytic structure of correlations functions has been seen to encode the flat space S-matrix~\cite{Maldacena:2011nz,Raju:2012zr,Arkani-Hamed:2015bza,Arkani-Hamed:2018kmz}.  In addition, these cosmological correlators display the analogues of simple poles and factorization associated with the exchange of new particles~\cite{Arkani-Hamed:2018bjr,Benincasa:2018ssx}.  These types of observations have helped demystify otherwise peculiar properties of these calculations.

In this work, we will explore the origin of secular divergences in de Sitter space and their relation to physics in flat space.  To make the origin of these divergences manifest, we will study theories that flow to perturbative IR fixed points in flat space.  At the fixed point, the theories are conformal and their de Sitter correlators are determined by a Weyl transformation~\cite{Farnsworth:2017tbz} from flat space to de Sitter, 
\beq\label{eq:conformalmap}
\langle \O_1(\x_1, \tau_1) .. \O_n(\x_n, \tau_n) \rangle_{\rm dS} = \left(\prod_{i =1}^n a(\tau_i)^{-\widetilde{\Delta}_i}\right) \langle \O_1(\x_1, \tau_1) .. \O_n(\x_n, \tau_n) \rangle_{\rm flat}  \ ,
\eeq
where $\widetilde{\Delta}_i$ are the dimensions of the operators at the IR fixed point, $\tau$ is the conformal time, $a(\tau)= (-H \tau)^{-1}$ is the scale factor and $\x$ are the spatial coordinates.  In perturbation theory, $\widetilde{\Delta}_i =\Delta_i + \gamma_i$ where $\Delta_i$ is the dimension at the UV fixed point and $\gamma_i$ is the anomalous dimension calculated in perturbation theory.  Expanding the dS correlator in $\gamma_i$, one sees that perturbation theory in de Sitter must contain $(\gamma_i \log a(\tau_i))^N$ divergences that are not present in the flat space calculation.  Given only the lowest order divergence, $\gamma_i \log a(\tau_i)$, one can use the DRG to recover the full power-law in Equation~(\ref{eq:conformalmap}) as required by conformal invariance.  

When the coupling is not tuned to be at the IR fixed point, the theory is not conformal and the de Sitter correlators are not necessarily related to flat space correlators by a Weyl transformation.  Yet, the perturbative calculation does not depend on the precise value of the coupling and the form of the secular divergences remains unchanged.  We will show that these logarithmic divergences can still be resummed and the resulting power law is determined by the (scale-dependent) flat space anomalous dimension, $\gamma(\mu)$, calculated at the scale of horizon crossing, $\mu=H$. 

We will first analyze a general version of this problem using conformal perturbation theory in de Sitter space.  When conformal perturbation theory is applicable in flat space, the same expansion can be used in de Sitter space using the map described in Equation~(\ref{eq:conformalmap}).  We show explicitly in conformal perturbation theory that anomalous dimensions in flat space become secular divergences in de Sitter.  We then show how they can be resummed using the DRG equations.  These insights will apply to a wide variety of theories, including perturbative QFTs involving gauge fields, fermions and/or conformally coupled scalars. We will show this explicitly in several examples.

The organization of this paper is as follows: in Section~\ref{sec:conformal}, we demonstrate our main results using conformal perturbation theory.   In Sections~\ref{sec:lambda} and~\ref{sec:yukawa}, we show how these general results arise in the specific examples of a conformally coupled scalar with a $\lambda \phi^4$ interaction in  $d=4-\epsilon$ dimensions and Yukawa interactions in four dimensions.  We conclude in Section~\ref{sec:conclusions}.  Details of the calculations are presented in the Appendices. 


\section{Conformal Perturbation Theory}\label{sec:conformal}

\subsection{Definition}
Conformal perturbation theory in flat space is a powerful tool for understanding a deformation around any fixed point, whether weakly or strongly coupled.  We imagine that the fixed point is described in terms of some action, $S_{\rm CFT}$, that is deformed by one of the operators in the CFT, 
\beq
S= S_{\rm CFT} + \lambda \mu^{d-\Delta} \int d^d x \, \O(\bx)\ , 
\eeq
where $\bx$ is a $d$-vector, $\mu$ is the renormalization scale, $\lambda$ is the dimensionless coupling, and $\Delta$ is the dimension of the operator.  From the Euclidean path integral description, it is easy to see that a correlation function in the perturbed theory can be related to a correlation function in the CFT via
\beq
\langle  \O_i(\by) \ldots  \rangle =  \langle \O_i(\by) \ldots e^{- \lambda \mu^{d-\Delta_j} \int d^d x \, \O_j(\bx) } \rangle_{\rm CFT} \ ,
\eeq
where $\langle ... \rangle_{\rm CFT}$ means we are calculating a correlation function in the (unperturbed) CFT.  Taylor expanding the exponential then gives the result in terms of correlation functions CFT,
\beq
\langle \O_i(\by) \ldots  \rangle =  \sum_n\frac{(- \lambda \mu^{d-\Delta_j})^n}{n!} \, \left(\prod_{k=1}^n \int d^d x_k \right) \langle  \O_i(\by)  \ldots \prod_{k=1}^n \O_j(\bx_k)  \rangle_{\rm CFT} \ .
\eeq
This procedure is very general and is even applicable to theories where $S_{\rm CFT}$ is unknown (or doesn't exist).  Of course, as a practical tool for calculations, it is limited to cases where the correlation functions are known and can be integrated.  

Given a theory in flat space described by conformal perturbation theory, we can apply the same procedure to define perturbation theory in de Sitter space, now writing $S= S_{\rm CFT} + \lambda \mu^{d-\Delta_j} \int d\tau d^{d-1} x \sqrt{-g} \, \O_j(\bx)$.  Using $ds^2 = a(\tau)^2 (-d\tau^2 + d\x^2)$ and $\sqrt{-g} = a^d(\tau)$ in conformal time, we can write 
\beq\label{eq:CPT_dS}
\langle \O_i(\by) \ldots  \rangle_{\rm dS} =   a^{- \Delta_i}(\tau_y)\sum_n\frac{(- \lambda \mu^{d-\Delta_j})^n}{n!} \left(\prod_{k=1}^n \int d^d x_k a^{d-\Delta_j} (\tau_k)\right) \langle  \O_i(\by)  \ldots \prod_{k=1}^n \O_j(\bx_k)  \rangle_{\rm CFT}
\eeq
where $\langle\ldots \rangle_{\rm CFT}$ is the correlation function at the UV fixed point in flat space.

\subsection{Perturbative Flow between Fixed Points}

We will consider a CFT in $d=4$ dimensions that contains an operator of dimension $\Delta = 4-\epsilon$.  We will first consider the RG flow in flat space and then address the behavior in de Sitter. In flat space, we deform the theory by $S = S_{\rm CFT} + \lambda \mu^{\epsilon} \int d^4 x \, \O(\bx)$.  

The first thing we must determine is what happens under RG flow.  At leading order, the $\beta$-function is given\footnote{$\mu$ is an energy scale which decreases to zero as we flow from the UV fixed point to the IR fixed point.} in terms of the dimension of the operator, $\beta_\lambda = \mu \frac{d \lambda}{d \mu} = -(4-\Delta)\lambda = -\epsilon \lambda$. We calculate the correction to the $\beta$ function, noticing that a generic correlator of quadratic order in $\lambda$ takes the form  
\beq
\langle \ldots \lambda^2 \mu^{2 \epsilon} \int d^4 x_1 d^4 x_2 \, \O(\bx_1) \O(\bx_2) \rangle \ .
\eeq
We will assume the operator product expansion (OPE) of $\O$ contains the term
\beq
\O(\bx_1) \O(\bx_2) \supset \frac{C}{|\bx_{12}|^{\Delta }} \O(\bx_2) \ ,
\eeq
where $\bx_{ij} = \bx_i-\bx_j$, and $C \neq 0$ is the OPE coefficient.  Using the OPE, our generic correlator contains the term
\begin{align}
    \nonumber
    \langle \ldots \frac{1}{2!} \lambda^2 \mu^{2 \epsilon} \int d^4 x_1 d^4 x_2 \, \O(\bx_1) \O(\bx_2) \rangle
    & \supset 
    \langle \ldots \frac{1}{2!} \lambda^2 \mu^{2 \epsilon} C \int d^4 x_{12}  \, |\bx_{12}|^{-4+ \epsilon} \int d^4 x_2 \, \O(\bx_2) \rangle \\
    \nonumber
    & \supset \langle \ldots \frac{1}{2!} \lambda^2 \mu^{2 \epsilon} C \, 2 \pi^2 \int_{0}^{\frac{1}{\mu}} \frac{d x_{12}}{|\bx_{12}|^{1 - \epsilon}} \int d^4 x_2 \, \O(\bx_2) \rangle \\[0.5em]
    & \supset \langle \ldots \frac{\lambda^2 \mu^{\epsilon} \pi^2 \, C}{\epsilon}  \int d^4 x_2 \, \O(\bx_2) \rangle \ ,
\end{align}
where $\mu$ is the normalization scale in units of energy.  We have regulated the divergent integral with a cutoff in position space for clarity, but we will otherwise use dimensional regularization throughout.  This divergence can be absorbed into a counterterm $\delta_\lambda = +\frac{\pi^2 C \lambda}{\epsilon}$ by changing the action to $S + (\lambda \mu^\epsilon)(1+\delta_\lambda) \int d^4 x \, \O(x)$. Differentiating this modified coupling constant with respect to $\mu$, we find
\begin{align}
    \label{eq:beta1}
    \beta_\lambda = -\epsilon \lambda + \pi^2 C \lambda^2 + O(\lambda^3) \ .
\end{align}
This beta function drives the theory from the UV fixed point at $\lambda = 0$ to the IR fixed point at $\lambda_{\rm IR} = \frac{\epsilon}{\pi^2 C}$.  Since $\epsilon \ll 1$, the behavior at the IR (UV) fixed point can be understood purely from perturbation theory around the UV (IR) fixed point.  
\vskip 10pt
\noindent {\bf Two-point function in flat space:} \hskip 5pt Let us now derive the perturbative correction to the equal time two point function in flat space, in anticipation of the cosmological calculation.  Expanding to first order in conformal perturbation theory, we have
\begin{align*}
    \braket{\O(\x_1, \tau_0) \O(\x_2, \tau_0) }
        = \langle \O(\x_1, \tau_0) \O(\x_2, \tau_0) \rangle _* - \lambda \, \mu^\epsilon \langle \O(\x_1, \tau_0) \O(\x_2, \tau_0) \int d^4 x_3 \O(\x_3) \rangle_* \ ,
\end{align*}
where $\langle \ldots \rangle_*$ is a correlation function calculated at the UV fixed point.  Inserting the correlation function (\ref{eq:dSthree}) with $a = 1$ and using (\ref{eq:radialautocorrelation}) to perform the integral, we find
\begin{align*}
    \langle \O(\x_1, \tau_0) \O(\x_2, \tau_0) \int d\tau_3 d^3 x_3 \, \O(\x_3, \tau_3)\rangle_* 
        = &\frac{C}{x_{12}^{\Delta}} \int \frac{d\tau_3 d^3 x_3}{|x_{13}^2 + \tau_E^2|^{\Delta/2} |x_{32}^2 + \tau_E^2|^{\Delta/2}} \\[1em]
        \overset{\epsilon \rightarrow 0}{\approx} \,
        &\frac{4 \pi^2 C}{ x_{12}^{2\Delta}} \left( \frac{1}{\epsilon} + \log (\mu x_{12}) + \dots \right) \ ,
\end{align*}
where $x_{ij} =|\x_{ij}|$. Reintroducing the coupling, $\lambda$, the bare two point correlation function is then
\begin{align*}
    \braket{\O(\x_1, \tau_0) \O(\x_2, \tau_0)}
        \approx \frac{1}{x_{12}^{2\Delta}} \left( 1 - \frac{4 \pi^2 C \lambda}{\epsilon} \, (\mu x_{12})^\epsilon \right)
\end{align*}
The $\frac{1}{\epsilon}$ divergence can be removed by introducing a counterterm $\delta_Z = - \frac{2 \pi^2 C \lambda}{\epsilon}$ so that the two-point function of the renormalized operator, $\O = Z \O_R$, takes the form
\begin{align}
    \braket{\O_R(\x_1, \tau_0) \O_R(\x_2, \tau_0))}
        \nonumber
        &= (1 - 2 \delta_Z) \braket{\O(\x_1, \tau_0) \O(\x_2, \tau_0))} \\
        &= \frac{1}{x_{12}^{2\Delta}} \left( 1 - 4 \pi^2 C \lambda \, \log(\mu x_{12}) + \dots \right)\ .
\end{align}
As a result, $\O_R$ acquires the anomalous dimension
\beq
    \gamma_\O = \mu \frac{d}{d \mu} (\delta_Z) = 2 \pi^2 C \lambda \ .
\eeq
Notice that at the IR fixed point the dimension of operator $\O$ becomes
\beq\label{eq:DeltaIR}
    \Delta_{\rm IR}
        = \Delta + \gamma_\O
        = 4 - \epsilon + 2 \pi^2 C \lambda_{\rm IR}
        = 4 + \epsilon \ ,
\eeq
which is consistent with the IR fixed point being attractive.  In fact, from the perspective of the IR fixed point, the RG flow from the deformation of $\lambda = \lambda_{\rm IR}+ \delta \lambda$ should be controlled by the dimension of the operator at the IR fixed point, 
\beq
\beta_{\delta \lambda} = -(4-\Delta_{\rm IR}) \delta \lambda +{\cal O}\left(\delta \lambda^2 \right) \ .
\eeq 
Taylor expanding Equation~(\ref{eq:beta1}) around the fixed point, one finds $\beta_{\delta \lambda} = \epsilon \delta \lambda$ as required from Equation~(\ref{eq:DeltaIR}). 
\vskip 10pt
\vskip 10pt
\noindent {\bf Two-point function in de Sitter space:} \hskip 5pt Now let us consider what happens to this theory when we compute late-time correlation functions in dS. At the UV fixed point, the power spectrum in dS follows from the conformal map from flat space to de Sitter,  Equation~(\ref{eq:conformalmap}),
\[
   \langle \O(\x_1,\tau_0) \O(\x_2, \tau_0)\rangle_{*,{\rm dS}} = \frac{a(\tau_0)^{-2\Delta}}{x_{12}^{2\Delta}} \ .
\]
We will use conformal perturbation theory in de Sitter, Equation~(\ref{eq:CPT_dS}), to determine the leading correction to this two point function.  From the outset, we know the theory in flat space flows to a CFT in the IR where $\O$ acquires a non-trivial anomalous dimension. Therefore, as explained in the introduction, there must be a $\log a(\tau_0)$ divergence associated with this two-point function in de Sitter space.  Our goal is to find this divergence explicitly and understand the behavior away from the fixed point.

Equal-time \textit{in-in} correlators are most easily computed using the analytic continuation to Euclidean time, as explained in Appendix~\ref{app:inin}.  Applying this formalism at linear order in $\lambda$ requires that we calculate the quantity
\begin{align}
    I   
        \nonumber
        = - \lambda \mu^\epsilon \int &d^3 x_3 \int_{-\infty}^{\infty} d \tau_E  \ a(i\tau_E + \tau_0)^4 \braket{\O(\x_1, \tau_0) \O(\x_2, \tau_0) \, \O(\x_3, i\tau_E + \tau_0)}_* \\[1em]
        \label{eq:dSleadingorderintegral}
        = - \lambda \mu^\epsilon C \, &\frac{a(\tau_0)^{-2\Delta}}{x_{12}^\Delta} \int_{-\infty}^{\infty} d \tau_E \, \int d^3 x_3 \, \frac{a(i\tau_E + \tau_0)^\epsilon}{|x_{23}^2 + \tau_E^2|^{\Delta/2}|x_{31}^2 + \tau_E^2|^{\Delta/2}} \ .
\end{align}
A priori, it might seem surprising that the above integral contains a divergence as $\tau_0 \to 0$.  At fixed $x_3 \neq x_1, x_2$, the integral in $\tau_E$ is manifestly convergent.  Similarly, at fixed $\tau_E$ the integral over $\x_3$ convergences.  However, we are integrating over both $\tau_E$ and $\x_3$ and there are divergences associated with taking $\x_3 \to \x_1, \x_2$ and $\tau_E \to 0$ simultaneously. We can estimate the degree of this divergence by noting that the integral of $\x_{3}$ around either $\x_{1}$ or $\x_2$ is regulated by $\tau_E$ so that $\int d^3 x_3 \approx \tau_E^3$.  If we then perform the $\tau_E$ integral, it scales as $\tau_E^{d- \Delta}= \tau^{\epsilon}_E$ which becomes a logarithmic divergence as $\epsilon \to 0$.  

The integral in Equation~(\ref{eq:dSleadingorderintegral}) is performed explicitly in Appendix~\ref{sec:conformaldSintegral} using Fourier transforms.  The resulting log-divergence arises precisely as expected from the above scaling argument, and leads to
\begin{align}
    I   
        \overset{(\ref{eq:maindSintegral})}{\approx}
        - \lambda C \, &\frac{a(\tau_0)^{-2\Delta}}{x_{12}^{2 \Delta}} \, 4 \pi^2
        \left( \frac{1}{\epsilon} + \log \left( -\frac{\mu x_{12}}{H \tau_0} \right) - \gamma_E + \dots \right) \ .
\end{align}
Putting it all together, the two point function of $\O$ is given by
\begin{eBox}
    \beq
        \braket{\O(\x_1, \tau_0) \O(\x_2, \tau_0)}_{\rm dS}
        = \frac{a(\tau_0)^{-2\Delta}}{x_{12}^{2\Delta}}
          \left( 1 - 4 \pi^2 C \lambda \left( \frac{1}{\epsilon} + \log \left( -\frac{\mu x_{12}}{H \tau_0} \right) + \dots \right) \right)
    \eeq
\end{eBox}
We see that the coefficient of the log is the anomalous dimension in flat space, $4 \pi^2 C \lambda = 2 \gamma_\O$.

\subsection{Dynamical RG}\label{subsec:DRG}

The two-point function in de Sitter space contains a number of divergent terms.  The first thing we should do is remove the $\frac{1}{\epsilon}$ divergence.  
Introducing a counterterm $\delta_Z = -\frac{4 \pi^2 C \, \lambda}{\epsilon}$, we get the renormalized two point function:
\begin{align*}
  \braket{\O_R(\x_1, \tau_0) \O_R(\x_2, \tau_0)}
  &= 
   \frac{a(\tau_0)^{-2\Delta}}{x_{12}^{2\Delta}}
    \left( 1 - 4 \pi^2 C \lambda \left( \frac{1}{\epsilon} + \log \left( -\frac{\mu x_{12}}{H \tau_0} \right) + \dots -\delta_Z \right) \right) \\[1em]
  &= \frac{a(\tau_0)^{-2\Delta}}{x_{12}^{2\Delta}}
    \left( 1 - 4 \pi^2 C \lambda \log \left( -\frac{\mu x_{12}}{H \tau_0} \right) + \dots \right) \ .
\end{align*}
The interpretation of the remaining log is more transparent if we separate the two dimensionless ratios as follows,
\beq
\log \left( -\frac{\mu x_{12}}{H \tau_0} \right) = \log \left(\frac{\mu }{H } \right)+\log \left( \frac{ x_{12}}{|\tau_0|} \right) \ .
\eeq
The distance must appear in the ratio $x_{12}/|\tau_0|$ in order to remain invariant under the rescaling $x \to \rho x$ and $a(\tau) \to \rho^{-1} a(\tau)$ which leaves the metric fixed.  Furthermore, additional interactions are known to give rise to pure $\log \mu /H$~\cite{Senatore:2009cf} and $\log x/\tau$ divergences and therefore must be treated separately.

We can easily eliminate $\log \mu /H$ divergences by choosing $\mu =H$.  Physically, this means that we should use standard RG to run the effective couplings of the theory to the energy scale $H$ (or simply define them at the scale $H$).  Since $H$ is fixed in de Sitter, this choice ensures that there will be no large logs associated with $\mu$.  We will define $\lambda_H = \lambda(\mu =H)$ as a reminder that we fixed the renormalization scale.  This is also a physically sensible result as $H$ is usually the physical scale where non-trivial (cosmological) correlations are generated.  

Renormalization in the conventional sense does not address the logarithmic growth in conformal time.  In fact, the mode is only super horizon when, $x_{12} / |\tau_0| \gg 1$,  and our log is necessarily large. We can formally resum the large logs in analogy with the renormalization group via the DRG.  Following the procedure in \cite{McDonald:2006hf,Burgess:2009bs}, we introduce a reference time and distance, $\tau_\star$ and $x_\star$.  We then add a counter-term to the operator, $\delta_Z \to \delta_Z (1+ 2\pi^2 C \lambda_H \log x_\star/|\tau_\star|)$, to get
\begin{align*}
  \braket{\O_R(\x_1, \tau_0) \O_R(\x_2, \tau_0)}
  &= \frac{a(\tau_0)^{-2\Delta}}{x_{12}^{2\Delta}}
    \left[ 1 - 4 \pi^2 C \lambda_H \log\left(\frac{x_{12} \tau_\star}{x_\star \tau_0} \right) \right] \ .
\end{align*}
Of course the operators of the theory are independent of $x_\star$ and $\tau_\star$ so that we have a differential equation for the two point function
\begin{align}
  \frac{\partial}{\partial \log(x_\star / |\tau_\star|)}\braket{\O_R(\x_1, \tau_0) \O_R(\x_2, \tau_0)}
    = 2\gamma(H) \braket{\O_R(\x_1, \tau_0) \O_R(\x_2, \tau_0)}
\end{align}
where we have defined
\beq
\gamma(H)  = \frac{\partial}{\partial \log(x_\star / |\tau_\star|) }\delta_Z|_{\mu = H} = 2 \pi^2 C \, \lambda_H + {\cal O}(\lambda_H^2) \ .
\eeq
Here we make a crucial assumption that all the secular terms can be absorbed with counter-terms such they vanish when $\tau_\star = \tau_0$ and $x_\star = x_{12}$.  We will return to discuss the justification for this assumption.

By construction, $x_\star / \tau_\star$ only appears in the ratio $x_{12}\tau_\star/(x_\star \tau_0)$ so we can rewrite this equation as 
\beq
\frac{\partial}{\partial \log(x_{12} / |\tau_0|)}\braket{\O_R(\x_1, \tau_0) \O_R(\x_2, \tau_0)}
    = - (2\Delta+ 2\gamma(H)) \braket{\O_R(\x_1, \tau_0) \O_R(\x_2, \tau_0)} \ ,
\eeq
where we introduced the additional factor of $2\Delta$ to account for the tree-level power spectrum.  We can solve this equation to find
\beq
\braket{\O_R(\x_1, \tau_0) \O_R(\x_2, \tau_0)} = \frac{1}{|a(\tau_0) x_{12}|^{2 \Delta + 2\gamma(H)}}\left(1+ O(\lambda_H)\right) \ .
\eeq
where the ${\cal O}(\lambda_H)$ corrections are not logarithmically enhanced.  We have used our freedom to choose the overall normalization of the operator to write the result in terms of $a(\tau_0 ) = -1/(H \tau_0)$.

An important open question we will not address in this work is the range of applicability of the DRG for cosmological correlators.  Instead, we will use the proximity to a conformal fixed point ensures that the DRG is accurately resuming our secular terms in the cases of interest.  At the conformal fixed point, the DRG will resum all the secular logs as required by symmetry.  Away from the fixed point, the structure of perturbation theory ensures the DRG will resum the leading logs as desired.  However, for a generic theory in de Sitter space, the applicability of the DRG is less certain.  In would be desirable to have a general result, like in flat space~\cite{Polchinski:1983gv}, that characterizes the validity and limitations of the DRG.  

\subsection{Summary}\label{subsec:summary}

Using conformal perturbation theory, we have seen that for a theory that flows between two fixed points, the (leading-log) two-point function in de Sitter space is given by 
\begin{eBox}
\beq\label{eq:DRGfinal}
\braket{\O_R(\x_1, \tau_0) \O_R(\x_2, \tau_0)} = \frac{1}{|a(\tau_0) x_{12}|^{2 \Delta + 2\gamma(H)}}\\ ,
\eeq
\end{eBox}
where $\gamma(H)= \gamma(\lambda(\mu = H))$ is the anomalous dimension calculated in flat space at the renormalization scale $\mu =H$.  Since the correlators in de Sitter are invariant under the group of de Sitter isometries, the higher point correlators must be de Sitter invariant with an effective scaling dimension of $\bar \Delta = \Delta + \gamma(H)$.  

The derivation of this result implicitly assumed that we were studying a deformation by a slightly relevant operator.  However, we must also find the same result from the perspective of the IR fixed point, in which case the deformation would have been slightly irrelevant.

These results will apply to a wide range of interacting theories, including massless particles with spin and conformally coupled scalars.  When these theories are perturbative, they are close to the Gaussian fixed point and therefore can be understood as being close to a CFT.  Light scalar fields in de Sitter are not captured by this description because Equation~(\ref{eq:conformalmap}) does not apply when $m^2/H^2 \neq d(d-2)/4$.


\section{Scalar Field Theory}\label{sec:lambda}

We would like to see how the general behavior described in Section~\ref{sec:conformal} arises in explicit examples.  In this section, we will calculate the power spectrum of the $\phi^2$ operator in the $\lambda \phi^4$ theory in dS.  These types of self-interactions are particularly common for inflationary models and are of broad interest. From flat space, we know an anomalous dimension arises at one-loop and  thus the same should be true in de Sitter.  Furthermore, by choosing $d=4- \epsilon$, the theory flows to the Wilson-Fisher fixed point and, as a consequence, the dynamics in de Sitter must approach the behavior of the CFT by construction.  

\subsection{The \texorpdfstring{$\braket{\phi^2 \phi^2}$}{phi squared} correlator}

Given a free real scalar field $\phi$ of mass $m$ in de Sitter space, one expands the fields in modes according to
\[
    \phi(\x,\tau) = \int \frac{d^{d-1} k}{(2 \pi)^{d-1}} e^{i \k \cdot \x} \
    \{ v_{\k}(\tau)a_{\k}+ v^*_{\k}(\tau)a^{\dagger}_{\k} \} \ .
\]
In the Bunch-Davies vacuum, the modes are given by
\[
    v_\k(\tau) = -i e^{i \left( \nu + \frac{1}{2} \right)\frac{\pi}{2}}\frac{\sqrt{\pi}}{2} H^{\frac{d-2}{2}} (-\tau)^\frac{d-1}{2} H_\nu(-\K \tau) \ ,
\]
where $k =|\k|$ and $H_\nu$ is the Hankel function of the first kind with $\nu = \sqrt{\frac{(d-1)^2}{4} - \frac{m^2}{H^2}}$. The free field theory in flat space maps to the conformal theory with mass $m^2 = d(d-2) H^2/4$ in dS, which means $\nu = \frac{1}{2}$. Therefore the mode functions are
\[
    v_\k(\tau) = \frac{-i}{a(\tau)^{\frac{d-2}{2}}} \frac{e^{-i \K \tau}}{\sqrt{2 \K}} \ .
\]
These are just the modes for a free massless theory in flat space, scaled by factors of $a^{-\Delta_\phi}$ as it should be in a conformal field theory.

As explained in Appendix~\ref{app:inin}, we can compute equal-time (\textit{in-in}) correlation functions as an anti-time-ordered Euclidean correlation function.  The anti-time-ordered propagator for our perturbative calculations is therefore
\begin{equation} \label{eq:propagator}
  \braket{\overline{T} \left( \phi(\k ,i\tau_E+\tau_0)\phi(-\k,\tau_0) \right)} = \frac{1}{a(i\tau_E+\tau_0)^{\Delta_\phi}a(\tau_0)^{\Delta_\phi}} \frac{e^{-\K|\tau_E|}}{2\K} \ .
\end{equation}
Note that $\tau_0 <0 $ will be fixed throughout the calculation and $\tau_E \in (-\infty,\infty)$.  

\begin{figure}
    \begin{subfigure}{.45\textwidth}
      \centering
      \includegraphics[scale=0.44]{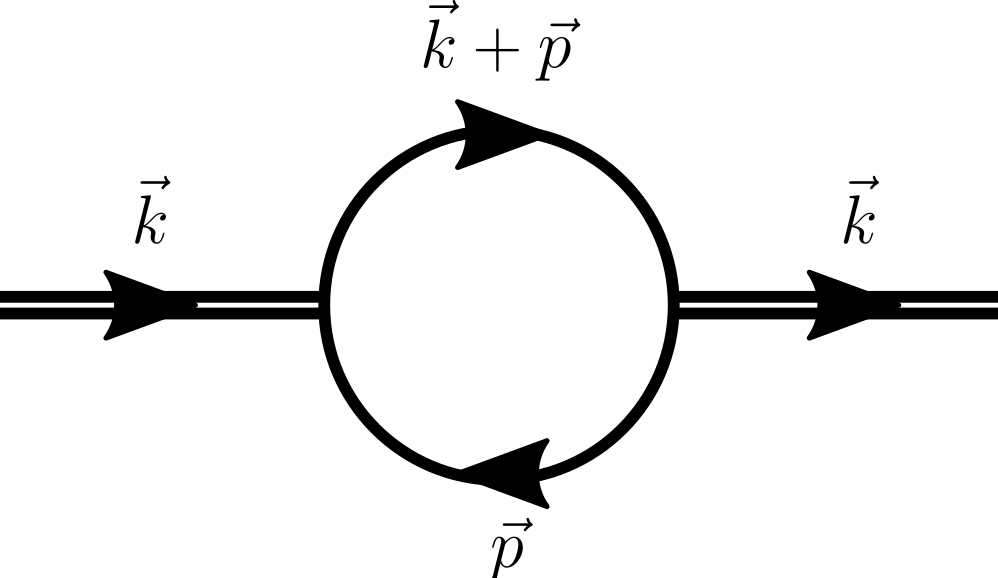}
      \caption{The tree level diagram}
      \label{fig:phi2phi2treelevel}
    \end{subfigure}
    \begin{subfigure}{.45\textwidth}
      \centering
      \includegraphics[scale=0.44]{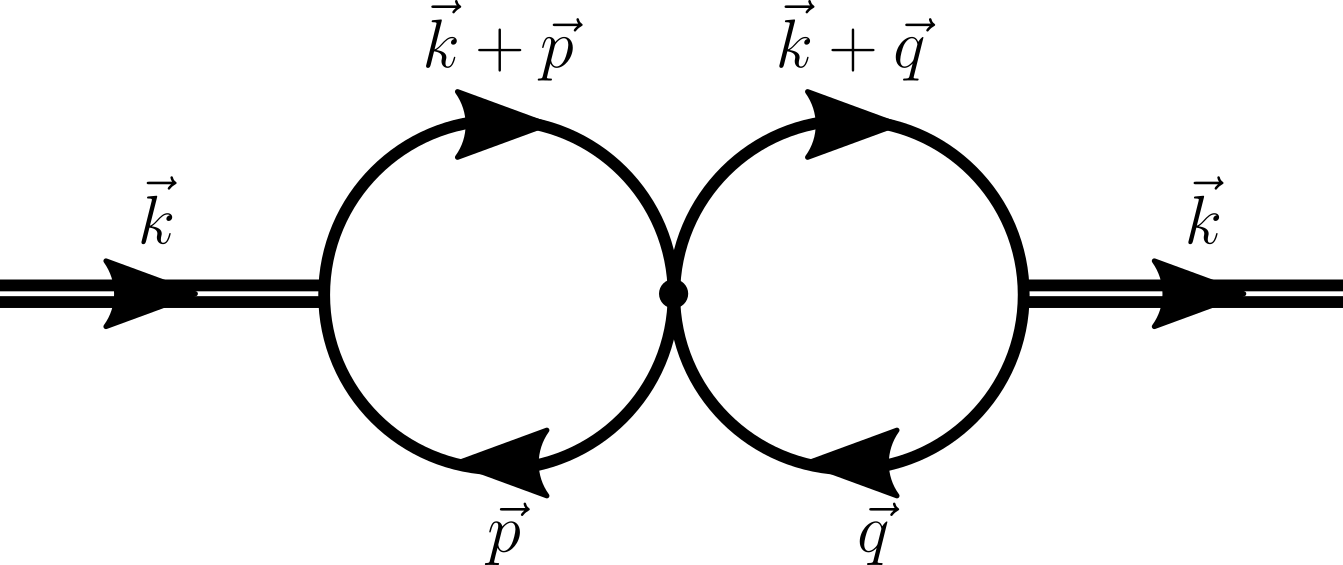}
      \caption{The order $\lambda$ correction}
      \label{fig:phi2phi2firstordercorr}
    \end{subfigure}
    \caption{The Feynman diagrams involved in the calculation of $\braket{\phi^2 \phi^2}$.}
\end{figure}

We begin by computing the equal-time \textit{in-in} correlator $\braket{\phi^2 \phi^2}$ in $d=4-\epsilon$ dimensions using the expression \cite{Green:2013rd}
\begin{equation} \label{eq:inincorrelator}
  \braket{\phi^2(\k,\tau_0)\phi^2(-\k,\tau_0)} = \braket{\overline{T} \left( \phi^2(\k,\tau_0)\phi^2(-\k,\tau_0) \exp[-\int_{-\infty}^{\infty} d \tau_E \ H_{\rm int}(i\tau_E+\tau_0)a(i\tau_E+\tau_0)] \right)} \ ,
\end{equation}
where $H_{\rm int}(\tau) = + \frac{\lambda \mu^\epsilon}{4!} \int d^{d-1} x \sqrt{-g} \phi^4(\tau,x)$. The lowest order term in this expansion corresponds to Fig. \ref{fig:phi2phi2treelevel} and it evaluates to
\begin{equation} \label{eq:phi2phi2treelevel}
  \braket{\phi^2(\k,\tau_0)\phi^2(-\k,\tau_0)}_*
  = \frac{1}{2 a^{4\Delta_\phi}(\tau_0)} \int \frac{d^D p}{(2 \pi)^D} \frac{1}{|\k + \p|p}
  = \frac{c}{2 a(\tau_0)^{2\Delta_{\phi^2}}} \ \K^{1-\epsilon} \ ,
\end{equation}
where $D = 3-\epsilon$, $d = D+1$ and $c \approx -1/(4\pi^2)$ is a constant (see (\ref{eq:prefactortreelevel})). Note that, $\Delta_\phi = \frac{d-2}{2}$ and $\Delta_{\phi^2} = d-2$. The computation of the momentum integral is detailed in Appendix~\ref{sec:appx_loopintegrals}. 

Now we can calculate the first order correction to $\braket{\phi^2 \phi^2}$ from the diagram in Figure~\ref{fig:phi2phi2firstordercorr}.  Expanding Equation~(\ref{eq:inincorrelator}) to first order gives
\begin{align}
    \nonumber
    -\frac{\lambda \mu^\epsilon}{4!} &\braket{\overline{T} \left(       \phi^2(\k,\tau_0)\phi^2(-\k,\tau_0) \int_{-\infty}^{\infty} d \tau_E \ a(\tau_3)^{D+1} \phi^4(\tau_3) \right)} \\[1em]
    \label{eq:firstorderintegral1}
    &= -\lambda \mu^\epsilon \int_{-\infty}^{\infty} d \tau_E \ a^d(\tau_3) \left( \int \frac{d^D p}{(2 \pi)^D}
    \frac{1}{a(\tau_3)^{2\Delta_\phi} a(\tau_0)^{2\Delta_\phi}} \frac{e^{-|\k+\p||\tau_E|}}{2|\k+\p|} \frac{e^{-p|\tau_E|}}{2\,p} \right)^2 \ .
\end{align}
We have used the shorthand $\tau_3 \equiv i\tau_E+\tau_0$ in the interest of space. The loop integral in the parentheses is computed in (\ref{eq:loopintegralsimplified}). Substituting this into (\ref{eq:firstorderintegral1}) the first order correction simplifies to
\beq \label{eq:eq:firstorderintegral2}
    -\lambda \mu^\epsilon \, \frac{\textcolor{black}{\K^{1-\epsilon}}}{a(\tau_0)^{4\Delta_\phi}} \, M^2
    \int_{-\infty}^{\infty} d \tau_E \ \frac{a(i\tau_E + \tau_0)^\epsilon}{|\tau_E|^{1-\epsilon}} \ K^2_{\frac{\epsilon-1}{2}}\left( \K |\tau_E| \right) \ .
\eeq
Comparing this with (\ref{eq:phi2phi2treelevel}) we see that tree level behavior of the correlation function has already factorized out.

Finally, we are left with only the time integral to compute. This integral is the source of the secular divergence. In the limit $|\k \tau_0| \ll 1$ and $\epsilon \rightarrow 0$, (\ref{eq:eq:firstorderintegral2}) is approximately
\beq
    +\frac{\lambda}{64\pi^4} \, \frac{ k^{1-\epsilon}}{a(\tau_0)^{4\Delta_\phi}}
    \left(
    \frac{1}{\epsilon} + \log \left(\frac{\mu}{H} \right) - \log ( - k \tau_0 ) + \dots
    \right) \ ,
\eeq
where $\dots$ are terms that vanish in the limit $|\k \tau_0| \ll 1$. The $\braket{\phi^2 \phi^2}$ correlation function at order $\lambda$ is then
    \begin{equation} \label{eq:phi2phi2fullcorrelator}
      \braket{\phi^2(\k,\tau_0)\phi^2(-\k,\tau_0)}
      = \frac{c}{2  a(\tau_0)^{2\Delta_{\phi^2}}} \ \K^{1-\epsilon}
      \left[
        1+\frac{\lambda}{32 \pi^4 c} \,
        \left(
        \frac{1}{\epsilon} + \log \left(\frac{\mu}{H} \right) - \log ( -\K \tau_0 ) + \dots
        \right) \ .
      \right]
    \end{equation}
Removing the divergence in (\ref{eq:phi2phi2fullcorrelator}) and performing a dynamic RG resummation:
\begin{align*}
    \braket{\phi^2(\k,\tau_0)\phi^2(-\k,\tau_0)}
    &\overset{\mu = H}{=}
    \frac{c}{2  a(\tau_0)^{2\Delta_{\phi^2}}} \ \K^{1-\epsilon}
    \exp \left(
      -\frac{\lambda_H}{32 \pi^4 c} \,
      \log ( -\K \tau_0 ) + \dots
    \right) (1+ \dots) \\
    &= \frac{c}{2  a(\tau_0)^{2\Delta_{\phi^2}}} \ \K^{1-\epsilon} ( -\K \tau_0 )^{2 \gamma_{\phi^2}(H)} \ (1+ O(\lambda_H^2)) \\
    &= \frac{c \, H^{-2\gamma_\phi(H)}}{2  a(\tau_0)^{2\Delta_{\phi^2} + 2 \gamma_{\phi^2}(H)}} \ \K^{1-\epsilon+2\gamma_{\phi^2}(H)} \ (1+ O(\lambda_H^2))
\end{align*}
where
\beq
\gamma_{\phi^2}(H) = -\frac{\lambda_H}{64 \pi^4 c} = +\frac{\lambda_H}{16 \pi^2} \ .
\eeq 
Comparing with (\ref{eq:phi2phi2treelevel}) we see that the effective dimension of the $\phi^2$ operator is corrected to:
\[
  \Delta_{\phi^2} \rightarrow \Delta_{\phi^2} + \gamma_{\phi^2}(H) = 2 - \epsilon + \frac{\lambda_H}{16 \pi^2}
\]
which is precisely $\Delta_{\phi^2} + \gamma_{\phi^2}(\mu =H)$ at one-loop, where $\gamma_{\phi^2}(\mu =H)$ is the anomalous dimension in flat space at the scale $\mu=H$.  
\subsection{Implications for \texorpdfstring{$\lambda \phi^4$}{lambda phi-4} in Four-Dimensions}\label{sec:lambdafourd}

As we discussed in Section~\ref{subsec:summary}, our results don't crucially require that there is an interacting IR fixed point.  As such, we can also view the above calculation as the dimensional regularization of $\lambda \phi^4$ in four-dimensions by taking the $\epsilon\to 0$ limit.  In that case, 
\beq
\Delta_{\phi^2}  = 2 + \frac{\lambda_H}{16\pi^2} 
\eeq
where $\lambda_H= \lambda(\mu =H)$ as before.  Unlike the $\epsilon > 0$ case, in four-dimensions the theory flows to the trivial fixed point in the IR, $\lambda(\mu \to 0) =0$.  Nevertheless, since the anomalous dimension is fixed at $\mu =H$ we still have a finite $\lambda_H$.  As a result, the power spectrum of $\phi^2$ in de Sitter space will acquire a fixed anomalous scaling with time and space in the super-horizon limit.

We can similarly conclude that conformally coupled scalars in de Sitter will acquire anomalous scaling in four-dimensions. The anomalous dimension for $\phi$ in $\lambda \phi^4$ is generated at two-loops and a direct calculation is beyond the scope of this work.  Nevertheless, we can conclude that such a two-loop de Sitter calculation should find that Equation~(\ref{eq:DRGfinal}) holds with 
\beq
\gamma_{\phi}(H) = \frac{\lambda_H^2}{12 (4 \pi)^4}  \ ,
\eeq
in accordance with the anomalous dimension in flat space.

\section{Yukawa Interaction}\label{sec:yukawa}

As a final example, we will study the RG flow of a scalar field theory in $d=4$ dimensions that is perturbed by a Yukawa coupling ${\cal L}_{\rm int} = \lambda \phi \bar \psi \psi$, where $\psi$ is a massless Dirac fermion.  In flat space, $\phi$ acquires an anomalous dimension at one loop and therefore will exhibit a 1-loop secular divergence in dS.  In this sense, the Yukawa coupling is the simplest example where the power spectrum of a fundamental scalar (as opposed to a composite operator) exhibits secular divergences of the type discussed in this paper.

The calculation of this effect does not require us to discuss the mode functions of the fermions directly.  For the purpose of our calculation, $\bar \psi \psi = \O$ is just an operator of dimension $\Delta = 3-\epsilon$ in a free CFT, where we have introduced $\epsilon$ as our regulator. We can therefore use the action
\beq
S[\phi] + \lambda \mu^\epsilon \int d^4 x \sqrt{|g|} \, \phi \, \O
\eeq
and evaluate the correlation functions of $\O$ using the same conformal perturbation theory approach described in Equation~(\ref{eq:CPT_dS}).  

The two point correlation function for $\phi$ receives a 1-loop ($O(\lambda^2)$) correction from the Yukawa interaction,
\begin{align}
 I \equiv \braket{\phi(\k, \tau_0) \phi(-\k, \tau_0)}_{1-{\rm loop}}
  = &\frac{(\lambda  \mu^\epsilon)^2}{2!}\int_{-\infty}^{\infty} d \tau_E \, a^4(\tau)\int_{-\infty}^{\infty} d \tau_E' \, a^4(\tau') \\ & \braket{\overline{T} \left(
          \phi(\k, \tau_0) \phi(-\k, \tau_0)
             \phi(-\k,\tau) \O(\k, \tau)
             \phi(\k,\tau') \O(-\k, \tau') 
    \right)} \ , \nonumber
\end{align}
where $\tau \equiv i\tau_E + \tau_0$ and $\tau' \equiv i\tau_E' + \tau_0$.  We evaluate this correlator by Fourier transforming the real-space correlation function of $\O$ given in Equation~(\ref{eq:conformalmap}).
Using (\ref{eq:yukawa2ptfnO}) and (\ref{eq:propagator}) in (\ref{eq:yukawacorrection1}), we find
\begin{align}
    \nonumber
    I &=\frac{(\lambda  \mu^\epsilon)^2}{a(\tau_0)^2} \ N \,
        \int_{-\infty}^{\infty} d \tau_E \, a^\epsilon(\tau)
        \int_{-\infty}^{\infty} d \tau_E' \, a^\epsilon(\tau')
             \int_\infty^\infty \frac{d\omega}{2 \pi} \frac{e^{-\K \left( |\tau_E|+|\tau_E'| \right) + i \omega (\tau_E - \tau_E')}}{(2 \K)^2} (k^2 + \omega^2)^{1-\epsilon} \\[1em]
    \label{eq:yukawafactorized}
    &=  \frac{(\lambda \mu^\epsilon)^2}{a(\tau_0)^2} N \,
        \int_\infty^\infty \frac{d\omega}{2 \pi} \
        \I (\omega,\K) \ \I (-\omega,\K) \, (k^2 + \omega^2)^{1-\epsilon} \ ,
\end{align}
where we have factorized the two time integrals by defining:
\beq \label{eq:yukawafactoredintegral}
    \I (\omega,\K) = \int_{-\infty}^{\infty} d \tau_E \, a^\epsilon(i\tau_E + \tau_0) \ \frac{e^{-\K |\tau_E| + i \omega \tau_E}}{2 \K} \ .
\eeq
We could compute this integral explicitly and substitute it back into (\ref{eq:yukawafactorized}) to proceed. However, the calculations become complicated if we take that route. We will therefore pursue a simpler strategy: first, we will evaluate (\ref{eq:yukawafactoredintegral}) in flat space-time i.e. we set $a(\tau) = 1$. Next, we will compute (\ref{eq:yukawafactoredintegral}) setting $\tau_0 = 0$. In each case we substitute the result back into (\ref{eq:yukawafactorized}) and extract the $\epsilon \rightarrow 0$ behavior. The final answer for the general $\tau_0 \neq 0$ case is then obtained by requiring that it match with these calculations in the respective limits. The details are given in Appendix \ref{sec:appx_yukawa}; the final results are (see (\ref{eq:yukawacorrectionflat}) and (\ref{eq:yukawacorrectiondS}))
\beq
    I   \overset{a = 1}{=}
        - \frac{ \lambda^2}{16 \pi^2 \, k} \ \left( \frac{1}{\epsilon} + 2 \log\left( \frac{\mu}{ k} \right) - \gamma_E + \dots \right)
\eeq
\beq \label{eq:yukawatau0zero}
    I   \overset{\tau_0 \rightarrow 0}{=}
        - \frac{ \lambda^2}{16\, \pi^2 \, a(\tau_0)^2 \, \K} \ \left( \frac{1}{\epsilon} + 2 \log\left( \frac{\mu}{H} \right) +  \mathcal{A} + \dots \right) \ ,
\eeq
where $\mathcal{A}$ is a divergent piece defined in (\ref{eq:divergentpiece}). Requiring that the general answer must reduce to these expressions, we find the first order correction (\ref{eq:yukawafactorized}) is
\begin{align}
    I   \overset{\epsilon, \tau_0 \rightarrow 0}{\approx}
        - \frac{\lambda^2}{16 \,\pi^2 \,  a(\tau_0)^2 \, \K} \ \left( \frac{1}{\epsilon} + 2 \log\left( -\frac{\mu}{\K H \tau_0} \right) + \dots \right) \ .
\end{align}
Note that setting $\tau_0 = 0$ in the $\log$ will cause it to blow up. This is the origin of the term $\mathcal{A}$ in (\ref{eq:yukawatau0zero}). Putting it all together, $\braket{\phi \phi}$ is
\begin{equation} \label{eq:yukawatwoptfn}
        \braket{\phi(\k, \tau_0) \phi(-\k, \tau_0)}
        = \frac{1}{2 \, a(\tau_0)^2 \, \K} \left(
          1 - \frac{\lambda^2}{8 \pi^2 } \left( \frac{1}{\epsilon} + 2 \log\left( -\frac{\mu}{\K H \tau_0} \right) + \dots \right)
          \right) \ .
    \end{equation}
 We remove the divergence in (\ref{eq:yukawatwoptfn}) using a counterterm $\delta_Z = -\frac{\pi^2 \lambda^2}{8 \, \epsilon}$ and perform a dynamical RG resummation to find
\begin{align*} \label{eq:yukawatwoptfn}
  \braket{\phi(\k, \tau_0) \phi(-\k, \tau_0)}
  &\overset{\mu = H}{=}  \frac{H^{-2\gamma_\phi(H)}}{2 \, a(\tau_0)^{2+2\gamma_\phi(H)} \, \K^{1-2\gamma_\phi(H)}} \ ,
\end{align*}
where 
\beq
\gamma_\phi(H) = +\frac{ \lambda_H^2}{8\pi^2} \ .
\eeq 
Comparing with (\ref{eq:propagator}) we see that the dimension of the $\phi^2$ operator is corrected to:
\[
    \Delta_\phi \rightarrow \Delta_\phi + \gamma_\phi(H) = 1 + \frac{ \lambda_H^2}{8 \pi^2} \ .
\]
As expected, $\gamma_\phi(H)$ is precisely the anomalous dimension found in four-dimensional flat space from a Yukawa coupling of a scalar to a Dirac fermion.

\section{Conclusions}\label{sec:conclusions}

Secular divergences present a significant challenge to perturbative calculations of cosmological correlators.  In this paper, we have shown that a certain class of such divergences have their origins as anomalous dimensions in flat space.  Like anomalous dimensions, they can be resummed to give corrections to the power law behavior at late times.  This interpretation of the divergences is unambiguous as this resummation is required to match the predictions at the conformal fixed point.

Our results apply to a wide range of quantum field theories in de Sitter space.  Massless particles with spin are generically conformally coupled and admit a description in terms of conformal perturbation theory.  In contrast, scalar fields with generic masses are not conformal in de Sitter and thus our treatment is limited to the case $m^2 \approx d(d-2) H^2/4$.  Scalar fields of generic masses are far from conformal and display a wider range of IR phenomena not addressed here.  Ultimately, one hopes to have a complete understanding of secular divergences both of quantum fields in de Sitter and in the presence of dynamical gravity.  

The main conclusion from this paper is that there is a broad class of secular divergences that is expected in QFT in de Sitter space and has an unambiguous and simple interpretation.  The meaning of IR divergences in de Sitter (particularly in the presence of gravity) has been the subject of significant ongoing interest and we believe these simple calculable examples can serve as a useful test-bed for future investigations.  

The broader problem of characterizing all possible IR divergences of cosmological correlators remains an outstanding problem.  Significant progress has been made recently on the divergences associated with massless scalars in de Sitter~\cite{Burgess:2015ajz,Gorbenko:2019rza,Baumgart:2019clc} and ultimately confirm the validity of the stochastic inflation framework~\cite{Starobinsky:1986fx} for understanding the non-trivial long distance behavior.  While light scalar fields have certainly presented a unique challenge to perturbative calculations, we have seen here that there are still secular terms associated with massive fields that arise as an interplay between the short-distance and late-time behavior.  A complete understanding of all such divergences at the same level as QFT in flat space would be a desirable outcome of a renewed focus on IR effects in de Sitter.

\paragraph{Acknowledgements}

We are grateful to Daniel Baumann, Matthew Baumgart, Tim Cohen, Victor Gorbenko, Enrico Pajer and Ben Wallisch for helpful discussions. D.G. would also like to thank the participants of the `Amplitudes meet Cosmology' workshop\footnote{\small \url{ https://www.simonsfoundation.org/event/amplitudes-meet-cosmology-2019/}} for useful conversations.
D.\,G.~is supported by the US~Department of Energy under grant no.~DE-SC0019035.

\appendix

\section{Analytic Continuation of the In-In Formalism}\label{app:inin}

Throughout the paper, we have integrated in Euclidean time to simplify time integration.  In this appendix, we will review this procedure following~\cite{Green:2013rd}.

We will use the in-in formalism to define equal-time correlations functions in the Bunch-Davies vacuum.  These are most easily calculated using the interaction picture operators $Q_{\mathrm{int}}$ and Hamiltonian $H_{\mathrm{int}}(t)$ via~\cite{Weinberg:2005vy}
\beq
\langle Q(\tau_0)\rangle=\left\langle\bar{T} \exp \left[i \int_{-\infty(1+i \epsilon)}^{\tau_{0}} H_{\mathrm{int}}(\tau) a(\tau) d \tau\right] Q_{\mathrm{int}}(\tau_0) T \exp \left[-i \int_{-\infty(1-i \epsilon)}^{\tau_{0}} H_{\mathrm{int}}(\tau) a(\tau) d \tau\right]\right\rangle \ .
\eeq
The Hamiltonian is given in terms of a Hamiltonian density, $\mathcal{H}_{\mathrm{int}}(\tau, x)$, via \beq
H_{\mathrm{int}}(\tau)=\int d^{3} x \, a^{3}(\tau) \mathcal{H}_{\mathrm{int}}(\tau, x) \ . 
\eeq
The deformation of the contour by a factor of $(1 \pm i \epsilon)$ defines the Bunch-Davies vacuum and ensures that the integrals converge as $\tau \to -\infty$.  We can make this convergence manifest by Wick rotating the contours on the left and right of the operator by $\tau \rightarrow \pm i \tau_{E}+\tau_{0}$~\cite{Behbahani:2012be,Green:2013rd}.  The resulting expression for the in-in correlators becomes an anti-time ordered integral
\beq
\langle Q(\tau_0)\rangle = \left\langle \bar{T} \left( Q_{\mathrm{int}}(\tau_0) \exp \left[-\int_{-\infty}^{\infty} H_{\mathrm{int}}\left(i \tau_{E}+\tau_{0}\right) a\left(i \tau_{E}+\tau_{0}\right) d \tau_{E}\right]\right) \right\rangle
\eeq
This is particularly useful for CFT correlators, where the anti-time order correlators in Euclidean time are given by
\bea\label{eq:dStwo}
    \langle {\cal O}(i\tau_E+ \tau_0, \x){\cal O}(i\tau_E' + \tau_0,\x') \rangle &=& \frac{a(i\tau_E + \tau_0)^{-\Delta} a(i\tau_E'+\tau_0)^{-\Delta}}{\left[(\tau_E-\tau_E')^2+( \x- \x')^2\right]^\Delta} \\ \label{eq:dSthree}
    \langle {\cal O}(\x_1,i\tau_1){\cal O}( \x_2, i\tau_2){\cal O}( \x_3,i\tau_3)\rangle &=& \frac{C\; a(i\tau_1)^{-\Delta}a(i\tau_2)^{-\Delta}a(i\tau_3)^{-\Delta}  }{|x_{12}^2+\tau_{12}^2|^{\Delta/2} | x_{23}^2+\tau_{23}^2|^{\Delta/2} | x_{31}^2+\tau_{31}^2|^{\Delta/2} } \ ,
\eea
where $\tau_{ij} = \tau_i -\tau_j$ and $\x_{ij} = \x_i - \x_j$.  We left the $\tau_0$ dependence in Equation~(\ref{eq:dSthree}) implicit in the interest of space.  It is important to notice that all correlators are calculated at equal Lorentzian time, $\tau_0$, and therefore the $\tau_0$ dependence cancels in $\tau_{ij}$, as shown explicitly in Equation~(\ref{eq:dStwo}).  

\section{Details of Loop Integration}

The explicit calculations of some of the loop integrals in the main text are performed in this appendix.  A number of these integrals are divergent but are made finite with dimensional regularization and/or analytic continuation.  We will therefore explain the regularization schemes first and then apply it to the integrals needed for the main text.

\subsection{Dimensional Regularization}

Throughout this paper, integrals are regulated by a parameter $\epsilon$ that controls the scaling behavior of the integral.  This may or may not be related to the dimension of space-time, as indicated in each section. In Section~\ref{sec:conformal}, we have general dimension $d$ and an operator of dimension $\Delta =d -\epsilon$, where $d$ and $\epsilon$ are independent parameters.  It is therefore useful to think of the $d$ and $\Delta$ dependence of these integrals as independent.

We will often be interested in $d$-dimensional radial Fourier transform of the correlation functions.  As a result, we often have to evaluate the following integral
\beq \label{eq:radialfourierxfrm}
     \frac{1}{|\bx|^{2\Delta}} = \pi^{\frac{d}{2}} \ 2^{d-2\Delta} \
     \frac{\Gamma(\frac{d}{2}-\Delta)}{\Gamma(\Delta)} \int \frac{d^d k}{(2 \pi)^d} e^{i \bk \cdot \bx} \frac{1}{|\bk|^{d-2\Delta}} \ .
\eeq
This integral is convergent for $2\Delta < d$.  We will not be in this regime, but we will define the integral at other values of $\Delta$ by analytic continuation of the above formula.  This is a standard  technique in QFT but is also commonly used as a definition of the Fourier transform.  This choice is justified because the divergent contributions we are neglecting are associated with $\delta$-functions in position space (contact terms) and therefore vanish when $x \neq 0$.

For conformal perturbation theory, one is often calculating an integral over a conformal three-point function.  Using Equation~(\ref{eq:radialfourierxfrm}) and the convolution theorem, it is straightforward to show that
\beq \label{eq:radialautocorrelation}
  \int \frac{d^d x_3}{(x_{13}^2)^\Delta (x_{32}^2)^\Delta}
    = \pi^{\frac{d}{2}} \left( \frac{\Gamma \left(\frac{d}{2}-\Delta \right)}{\Gamma(\Delta)} \right)^2 \frac{\Gamma \left(2\Delta - \frac{d}{2} \right)}{\Gamma (d-2\Delta)} \, \frac{1}{x_{12}^{4\Delta-d}} \ .
\eeq
This result is again defined by analytic continuation, the divergent contributions are contact terms and vanish when $x_{12} \neq 0$.  

\subsection{de Sitter Conformal Perturbation Theory}
\label{sec:conformaldSintegral}

The leading correction to the two point correlation function $\braket{\O \O}$ in de Sitter space involves the integral (see (\ref{eq:dSleadingorderintegral}))
\[
    I = - \lambda \mu^\epsilon C \, \frac{a(\tau_0)^{-2\Delta}}{x_{12}^\Delta}
            \int_{-\infty}^{\infty} d \tau_E \, \int d^3 x_3 \, \frac{a(i\tau_E + \tau_0)^\epsilon}{|x_{23}^2 + \tau_E^2|^{\Delta/2}|x_{31}^2 + \tau_E^2|^{\Delta/2}} \ .
\]
Shifting $\x_3 \rightarrow \x_3 + \x_1$, we see that the $\x_3$ integral is just a convolution of the function $F(\x, \tau_E) = (x^2 + \tau_E^2)^{-\Delta/2}$ with itself,
\begin{align*}
  I &= - \lambda \mu^\epsilon C \, \frac{a(\tau_0)^{-2\Delta}}{x_{12}^\Delta}
          \int_{-\infty}^{\infty} d \tau_E \, a(i\tau_E + \tau_0)^\epsilon \int d^3 x_3 \, \frac{1}{|(\x_{21} - \x_3)^2 + \tau_E^2|^{\Delta/2}|x_3^2 + \tau_E^2|^{\Delta/2}} \\[1em]
  &= - \lambda \mu^\epsilon C \, \frac{a(\tau_0)^{-2\Delta}}{x_{12}^\Delta}
          \int_{-\infty}^{\infty} d \tau_E \, a(i\tau_E + \tau_0)^\epsilon \int \frac{d^3 k}{(2 \pi)^3} \, e^{i\k \cdot \x_{21}} \tilde{F}(\k, \tau_E)^2 \ .
\end{align*}
In the last step, we have applied the convolution theorem, 
\beq \label{eq:convolutiontheoremx}
    \int d^D y \ F(\x-\y) F(\y) = \int \frac{d^D k}{(2 \pi)^D} \ e^{i\k.\x} \tilde{F}(\k)^2 \ ,
\eeq
and introduced
\begin{equation*}
  \tilde{F}(\k, \tau_E)
  = \int d^3 x \, e^{-i\k \cdot \x} \, \frac{1}{(\X^2 + \tau_E^2)^{\Delta/2}}
  = \frac{2 \pi^{3/2}}{\Gamma \left(2-\frac{\epsilon}{2}\right)} \left(\frac{\K}{2\,|\tau_E|}\right)^{\frac{1-\epsilon}{2}} K_{\frac{\epsilon -1}{2}}( \K \left| \tau_E \right| ) \ . 
\end{equation*}
The full expression is therefore
\begin{align*}
  I &= - \lambda \mu^\epsilon C \, \frac{a(\tau_0)^{-2\Delta}}{x_{12}^\Delta}
        \int_{-\infty}^{\infty} d \tau_E  \, a(i\tau_E + \tau_0)^\epsilon
          \\
          &\qquad \quad \times \int \frac{d^3 k}{(2 \pi)^3} \, e^{i \k \cdot \x_{21}} \frac{4 \pi^3}{\Gamma \left(2-\frac{\epsilon}{2}\right)^2} \left(\frac{\K}{2\,|\tau_E|}\right)^{1-\epsilon} \left(  K_{\frac{\epsilon -1}{2}}(\K \left| \tau_E \right| ) \right)^2 \ .
\end{align*}
Computing the $k$ integral,
\begin{align*}
  I = - \lambda \mu^\epsilon C \, \frac{a(\tau_0)^{-2\Delta}}{x_{12}^\Delta} \
  & \frac{4 \pi^2}{\Gamma \left(2-\frac{\epsilon}{2}\right)^2}
  \int_{-\infty}^{\infty} d \tau_E  \, a(i\tau_E + \tau_0)^\epsilon \\
    &\times
    2^{-5+\epsilon} \, |\tau_E|^{-5 + 2\epsilon} \, \Gamma \left( \frac{5}{2}-\epsilon \right) \Gamma \left( 2-\frac{\epsilon}{2} \right) \,
    _2\tilde{F}_1\left(\frac{5}{2}-\epsilon ,2-\frac{\epsilon }{2};\frac{5}{2}-\frac{\epsilon }{2};-\frac{x_{12}^2}{4 \tau_E^2}\right) \ .
\end{align*}
where $_2\tilde{F}_1$ is the regularized hypergeometric function. Carrying out the integral over $\tau_E$ and taking the limits $\epsilon \rightarrow 0$ and $\tau_0 \rightarrow 0$, we get
\beq \label{eq:maindSintegral}
  I \approx - \lambda C \, \frac{a(\tau_0)^{-2\Delta}}{x_{12}^{2 \Delta}} \, 4 \pi^2 
    \left( \frac{1}{\epsilon} + \log \left( -\frac{\mu x_{12}}{H \tau_0} \right) - \gamma_E + \dots \right) \ .
\eeq

\subsection{Integrals in the \texorpdfstring{$\lambda \phi^4$}{lambda phi-4} theory}
\label{sec:appx_loopintegrals}

The equal time correlation function for the $\phi^2$ operator (see Equation~(\ref{eq:phi2phi2treelevel}) and Fig.~\ref{fig:phi2phi2treelevel}) involves the loop integral
\beq \label{eq:treelevelloopintegral}
    \int \frac{d^D p}{(2 \pi)^D} \frac{1}{|\k + \p|\P} \ .
\eeq
Usually, one computes loop integrals like this with Feynman parameters. 
However, there is another way to do this calculation which is particularly useful for unequal times. We notice that the integral (\ref{eq:treelevelloopintegral}) is just a convolution of the function $\K^{-1}$ with itself. Therefore, we can use the convolution theorem in the form
\beq \label{eq:convolutiontheoremp}
    \int \frac{d^D p}{(2 \pi)^D} \ \tilde{F}(\k-\p) \tilde{F}(\p) = \int d^D x \ e^{-i\k.\x} F(\x)^2 \ ,
\eeq
i.e. $\tilde{F} \ast \tilde{F} \overset{F.T}{\longleftrightarrow} F^2$. If $\tilde{F}(\p)$ is radially symmetric we can change $\p \rightarrow -\p$ without changing the result. Therefore, all we need to do is find the Fourier transform of $\K^{-1}$ and square it. Using (\ref{eq:radialfourierxfrm}) we see that $2\pi^2 \K^{-1} \overset{F.T}{\longleftrightarrow} ~ \ \X^{-2+\epsilon}$ and therefore
\[
  \int \frac{d^{3-\epsilon} k}{(2 \pi)^{3-\epsilon}} \frac{1}{|\k + \p|\P}
    \overset{(\ref{eq:convolutiontheoremp})}{=}
      \int d^{3-\epsilon} x \ e^{-i\k \cdot \x} \left( \frac{1}{2\pi^2 \X^{2-\epsilon}} \right)^2
    \overset{(\ref{eq:radialfourierxfrm})}{=}
      c \, \K^{1-\epsilon} \ ,
\]
where
\beq \label{eq:prefactortreelevel}
    c = \frac{\Gamma \left(1-\frac{\epsilon}{2}\right)^2 \Gamma \left(\frac{\epsilon-1}{2}\right)}{2^{3-\epsilon} \pi^{\frac{5-\epsilon}{2}} \Gamma(2-\epsilon )} \approx -\frac{1}{4 \pi^2} + O(\epsilon) \ .
\eeq
We can use the same strategy to evaluate the loop integral in (\ref{eq:firstorderintegral1}),
\beq \label{eq:loopintegralatdifftimes}
    \int \frac{d^{3-\epsilon} p}{(2 \pi)^{3-\epsilon}} \frac{e^{-|\k+\p||\tau_E|}}{2|\k+\p|} \frac{e^{-\P|\tau_E|}}{2\P} \ .
\eeq
Notice that the loop integral is a convolution of the function $\tilde{F}(\k) = \frac{e^{-\K |\tau_E|}}{2 \K}$ with itself. Moreover
\[
    \frac{e^{-\K|\tau_E|}}{2 \K}
        = \int_{-\infty}^{\infty} \frac{d\omega}{2 \pi} e^{i \omega \tau_E} \frac{1}{\omega^2 + k^2} \equiv \tilde{F}(\k) \ ,
\]
and therefore the Fourier transform of the function $\tilde{F}(\k)$ is
\begin{align*}
    F(\x) = \int \frac{d^D k}{(2 \pi)^D} \frac{e^{-\K|\tau_E| + i \k \cdot \x}}{2 \K}
    & = \int \frac{d^D k}{(2 \pi)^D} \int \frac{d\omega}{2 \pi} e^{i \omega \tau_E + i \k \cdot \x} \frac{1}{\omega^2 + \K^2} \\[1em]
    &\overset{(\ref{eq:radialfourierxfrm})}{=} \frac{\Gamma(\frac{2-\epsilon}{2})}{\pi^{\frac{4-\epsilon}{2}} 2^2} \ \ \frac{1}{(\X^2 + \tau_E^2)^\frac{2-\epsilon}{2}} \ .
\end{align*}
We can now apply the convolution theorem (\ref{eq:convolutiontheoremp}) to the loop integral in ({\ref{eq:loopintegralatdifftimes}}) to find
\beq \label{eq:loopintegralintermediate}
  \int \frac{d^{3-\epsilon} p}{(2 \pi)^{3-\epsilon}} \frac{e^{-|\k+\p||\tau_E|}}{2|\k+\p|} \frac{e^{-\P|\tau_E|}}{2\P}
    = 
    \frac{\Gamma^2(\frac{2-\epsilon}{2})}{\pi^{4-\epsilon} 2^4}
    \int d^{3-\epsilon} x \ \frac{e^{- i \k.\x}}{(\X^2 + \tau_E^2)^{2-\epsilon}} \ .
\eeq
This result can be simplified using (\ref{eq:radialfourierxfrm}) if we turn $d^D x \, e^{- i \k \cdot \x} \rightarrow d^d x \, e^{- i \bK \cdot \bx}$ where $\bK \equiv (\omega, \k)$ and $\bx \equiv (\tau, \x)$. To accomplish this we rewrite
\[
  \frac{1}{(\X^2 + \tau_E^2)^{2-\epsilon}}
    = \int_{-\infty}^{\infty} d \tau \, \delta (\tau-\tau_E) \frac{1}{(\X^2 + \tau^2)^{2-\epsilon}}
    = \int_{-\infty}^{\infty} d \tau \, \int_{-\infty}^{\infty} d \omega \, e^{-i\omega(\tau-\tau_E)} \frac{1}{(\X^2 + \tau^2)^{2-\epsilon}} \ , 
\]
so that our integral becomes
\begin{align*}
    \int d^{3-\epsilon} x \ \frac{e^{- i \k.\x}}{(\X^2 + \tau_E^2)^{2-\epsilon}}
    &= \int_{-\infty}^{\infty} d \omega \, e^{i\omega \tau_E}
    \int d^{4-\epsilon} x \ \frac{e^{- i \bK \cdot \bx}}{|\bx|^{4-2\epsilon}} \\[1em]
    &\overset{(\ref{eq:radialfourierxfrm})}{=}
    \pi^{\frac{4-\epsilon}{2}} \ 2^{\epsilon} \ \frac{\Gamma(\frac{\epsilon}{2})}{\Gamma(2-\epsilon)}
    \int_{-\infty}^{\infty} d \omega \, \frac{e^{i\omega \tau_E}}{(\omega^2+\K^2)^{\epsilon/2}} \\[1em]
    &= \frac{\pi^{\frac{3-\epsilon}{2}} \ 2^{\frac{1+\epsilon }{2}}}{\Gamma(2-\epsilon)}
    \left(\frac{\K}{|\tau_E|}\right)^{\frac{1-\epsilon }{2}}
    K_{\frac{\epsilon-1}{2}} \left( \K |\tau_E| \right) \ ,
\end{align*}
where $K_\nu(z)$ is the modified Bessel function of the second kind. Plugging this back into Equation~(\ref{eq:loopintegralintermediate}), the loop integral turns out to be
\beq \label{eq:loopintegralsimplified}
    \int \frac{d^{3-\epsilon} p}{(2 \pi)^{3-\epsilon}} \frac{e^{-|\k+\p||\tau_E|}}{2|\k+\p|} \frac{e^{-\P|\tau_E|}}{2\P}
    = M \left(\frac{\K}{|\tau_E|}\right)^{\frac{1-\epsilon }{2}}
    K_{\frac{\epsilon-1}{2}} \left( \K |\tau_E| \right)
\eeq
where
\beq
    M = \frac{\Gamma^2(\frac{2-\epsilon}{2})}
        {2^{\frac{7-\epsilon}{2}} \pi^{\frac{5-\epsilon}{2}} \Gamma(2-\epsilon)}
        \overset{\epsilon \rightarrow 0}{=}
        \frac{1}{\sqrt{128 \pi^5}} + O(\epsilon) \ .
\eeq

\subsection{Integrals for the Yukawa calculation}
\label{sec:appx_yukawa}

In the main text, we found the one-loop correction to the $\phi$ power spectrum was determined by the  correlation function
\begin{align} \label{eq:yukawacorrection1}
    I = (\lambda \mu^\epsilon)^2
        \int_{-\infty}^{\infty} d \tau_E \,  &a^4(\tau)
        \int_{-\infty}^{\infty} d \tau_E' \, a^4(\tau') \\
        & \,
        \braket{\overline{T} \left(
        \wick{\c1 \phi(\k, \tau_0) \c2 \phi(-\k, \tau_0) \c1 \phi(-\k,\tau) \c3 \O(\k, \tau) \c2 \phi(\k,\tau') \c3 \O(-\k, \tau')}
        \right)} \ , \nonumber
\end{align}
where we have shown the contractions required for a connected correlator.  To evaluate this correlation function we need the anti-time-ordered two-point correlation function of $\O$. However, unlike a generic operator in a CFT, the normalization of $\O = \bar\psi \psi$ is determined by the propagator of the free fermion.  With this normalization factor, the power spectrum of this operator is given by
\begin{align}
    \braket{\overline{T}(\O(\k, \tau) \O(-\k, \tau'))}
        &= \int d^3 x \ e^{-i \k \cdot \x} \braket{\overline{T}(\O(x, \tau) \O(0, \tau'))} \\
        &= \frac{1}{\pi^4}\int d^3 x \ e^{-i \k \cdot \x} \frac{a^{-\Delta}(\tau)a^{-\Delta}(\tau')}{(x^2 + (\tau_E -\tau_E')^2)^{3-\epsilon}} \\
        &\overset{(\ref{eq:radialfourierxfrm})}{=}
        \label{eq:yukawa2ptfnO}
        N \int_{-\infty}^{\infty} \frac{d \omega}{2 \pi} \ e^{i \omega (\tau_E - \tau_E')} (\K^2+\omega^2)^{1-\epsilon} a(\tau)^{-3+\epsilon}a(\tau')^{-3+\epsilon} \ ,
\end{align}
where, in the second line, the factor of $\pi^{-4}$ arises for matching to a Dirac fermion and where we have defined
\[
    N = \frac{1}{2^{2-2\epsilon} \pi^2} \ \frac{\Gamma(-1+\epsilon)}{\Gamma(3-\epsilon)}
       \overset{\epsilon \rightarrow 0}{=}
        - \frac{1}{8 \pi} \left( \frac{1}{\epsilon} - 2 \gamma_E + \dots \right) + O(\epsilon) \ .
\]
We now turn our attention to computing (\ref{eq:yukawafactorized}) in two simple cases: (i) In flat space-time and (ii) when $\tau_0 = 0$. Starting with the flat space-time limit, 
\[
    \I (\omega,\K)
    \overset{a(\tau) \rightarrow 1}{=}
    \int_{-\infty}^{\infty} d \tau_E \, \frac{e^{-\K |\tau_E| + i \omega \tau_E}}{2 \K}
    = \frac{1}{\K^2 + \omega^2} \ ,
\]
the first order correction (\ref{eq:yukawacorrection1}) is
\begin{align}
    \nonumber
    I_{\text{flat}} = (\lambda \mu^\epsilon)^2  N \,
        \int_\infty^\infty \frac{d\omega}{2 \pi} \ \frac{1}{\left( \K^2 + \omega^2 \right)^{1+\epsilon}}
    \label{eq:yukawacorrectionflat}
    \overset{\epsilon \rightarrow 0}{\approx}
        - \frac{ \lambda^2}{16 \, \pi^2 \, \K} \ \left( \frac{1}{\epsilon} + 2 \log\left( \frac{\mu}{\K} \right) - \gamma_E + \dots \right) \ .
\end{align}
Next, we return to de Sitter space but set $\tau_0 = 0$ to find
\begin{align}
    \nonumber
    \I (\omega,\K)
        &= \int_{-\infty}^{\infty} d \tau_E \, \left( -\frac{1}{H \, \tau_E} \right)^\epsilon \ \frac{e^{-\K |\tau_E| + i \omega \tau_E}}{2 \K} \\[1em]
        &= \frac{\Gamma(1-\epsilon)}{H^\epsilon} \ \frac{ i^\epsilon (\K-i\omega)^{1-\epsilon} + (-i)^\epsilon (\K+i\omega)^{1-\epsilon}}{2 \K \left( \K^2+\omega^2 \right)^{1-\epsilon}} \ .
\end{align}
This expression can be simplified using $\exp \left( {2 i \tan^{-1}\left( \frac{\omega}{\K} \right)} \right) = \frac{\K + i \omega}{\K - i \omega}$. Then, the integrand in (\ref{eq:yukawafactorized}) becomes
\begin{align}
    \nonumber
    \I (\omega,\K) \ &\I (-\omega,\K) \, (\K^2 + \omega^2)^{1-\epsilon}\\[1em]
    \nonumber
    &= \frac{\Gamma(1-\epsilon)^2}{H^{2\epsilon} \, \K^2}
          \cos \left( (1-\epsilon) \tan^{-1}\left( \frac{\omega}{\K} \right) + \frac{\pi \epsilon}{2} \right)
          \cos \left( (1-\epsilon) \tan^{-1}\left( \frac{\omega}{\K} \right) - \frac{\pi \epsilon}{2} \right) \\[1em]
    \label{eq:yukawaintegranddS}
    &\overset{\epsilon \rightarrow 0}{\approx}
       \frac{1}{\K^2 + \omega^2} \ \left( 1 + \left( 2 \, \gamma_E - 2 \log H + 2 \, \frac{\omega}{\K} \tan^{-1}\left( \frac{\omega}{\K} \right) \right) \epsilon + O(\epsilon^2) \right) \ .
\end{align}
We can now substitute (\ref{eq:yukawaintegranddS}) into (\ref{eq:yukawafactorized}) to obtain
\begin{align}
    \nonumber
    I_0 =
    \frac{(\lambda \mu^\epsilon)^2}{a(\tau_0)^2} N \,
      &\int_{\infty}^\infty \frac{d \omega}{2 \pi} \
        \frac{1}{\K^2 + \omega^2} \, \left( 1 + \left( 2 \, \gamma_E - 2 \log H + 2 \, \frac{\omega}{\K} \tan^{-1}\left( \frac{\omega}{\K} \right) \right) \epsilon + O(\epsilon)^2 \right) \\[1.5em]
    \label{eq:yukawacorrectiondS}
    &\overset{\epsilon \rightarrow 0}{\approx}
    - \frac{ \lambda^2}{16 \, \pi^2 \,  a(\tau_0)^2 \, \K} \ \left( \frac{1}{\epsilon} + 2 \log\left( \frac{\mu}{H} \right) +  \mathcal{A} + \dots \right) \ .
\end{align}
In the final step we have used
\[
  \int_{-\infty}^{\infty} \frac{d \omega}{2 \pi} \ \frac{1}{\K^2 + \omega^2} = \frac{1}{2 \K}
\]
and introduced
\beq \label{eq:divergentpiece}
  \frac{\mathcal{A}}{\K}
    = \int_{-\infty}^{\infty} \frac{d \omega}{2 \pi} \ \frac{\omega}{\K} \ \frac{\tan^{-1}(\omega/\K)}{\K^2 + \omega^2}
    = \frac{1}{2 \pi \K} \int_{-\pi/2}^{\pi/2} d \theta \, \theta \tan \theta
\eeq
with $\theta = \tan^{-1}\left( \frac{\omega}{\K} \right)$. This term is clearly divergent and, as pointed out in the main text, it is a consequence of setting $\tau_0 = 0$.

\clearpage
\phantomsection
\addcontentsline{toc}{section}{References}
\bibliographystyle{utphys}
\bibliography{dSRefs}

\providecommand{\href}[2]{#2}\begingroup\raggedright\begin{thebibliography}{10}

\bibitem{Bousso:1999cb}
R.~Bousso, ``{Holography in general space-times},''
  \href{http://dx.doi.org/10.1088/1126-6708/1999/06/028}{{\em JHEP} {\bfseries
  06} (1999) 028},
\href{http://arxiv.org/abs/hep-th/9906022}{{\ttfamily arXiv:hep-th/9906022
  [hep-th]}}.

\bibitem{Strominger:2001pn}
A.~Strominger, ``{The dS / CFT correspondence},''
  \href{http://dx.doi.org/10.1088/1126-6708/2001/10/034}{{\em JHEP} {\bfseries
  10} (2001) 034},
\href{http://arxiv.org/abs/hep-th/0106113}{{\ttfamily arXiv:hep-th/0106113
  [hep-th]}}.

\bibitem{Witten:2001kn}
E.~Witten, ``{Quantum gravity in de Sitter space},'' in {\em {Strings 2001:
  International Conference Mumbai, India, January 5-10, 2001}}.
\newblock 2001.
\newblock
\href{http://arxiv.org/abs/hep-th/0106109}{{\ttfamily arXiv:hep-th/0106109
  [hep-th]}}.
\newblock

\bibitem{Mazur:2001aa}
P.~O. Mazur and E.~Mottola, ``{Weyl cohomology and the effective action for
  conformal anomalies},''
  \href{http://dx.doi.org/10.1103/PhysRevD.64.104022}{{\em Phys. Rev.}
  {\bfseries D64} (2001) 104022},
\href{http://arxiv.org/abs/hep-th/0106151}{{\ttfamily arXiv:hep-th/0106151
  [hep-th]}}.

\bibitem{Maldacena:2002vr}
J.~M. Maldacena, ``{Non-Gaussian features of primordial fluctuations in single
  field inflationary models},''
  \href{http://dx.doi.org/10.1088/1126-6708/2003/05/013}{{\em JHEP} {\bfseries
  05} (2003) 013},
\href{http://arxiv.org/abs/astro-ph/0210603}{{\ttfamily arXiv:astro-ph/0210603
  [astro-ph]}}.

\bibitem{Alishahiha:2004md}
M.~Alishahiha, A.~Karch, E.~Silverstein, and D.~Tong, ``{The dS/dS
  correspondence},'' \href{http://dx.doi.org/10.1063/1.1848341}{{\em AIP Conf.
  Proc.} {\bfseries 743} no.~1, (2004) 393--409},
\href{http://arxiv.org/abs/hep-th/0407125}{{\ttfamily arXiv:hep-th/0407125
  [hep-th]}}.

\bibitem{Ford:1984hs}
L.~H. Ford, ``{Quantum Instability of De Sitter Space-time},''
\href{http://dx.doi.org/10.1103/PhysRevD.31.710}{{\em Phys. Rev.} {\bfseries
  D31} (1985) 710}.

\bibitem{Antoniadis:1985pj}
I.~Antoniadis, J.~Iliopoulos, and T.~N. Tomaras, ``{Quantum Instability of De
  Sitter Space},''
\href{http://dx.doi.org/10.1103/PhysRevLett.56.1319}{{\em Phys. Rev. Lett.}
  {\bfseries 56} (1986) 1319}.

\bibitem{Starobinsky:1986fx}
A.~A. Starobinsky, ``{Stochastic de Sitter (Inflationary) Stage in the Early
  Universe},''
\href{http://dx.doi.org/10.1007/3-540-16452-9_6}{{\em Lect. Notes Phys.}
  {\bfseries 246} (1986) 107--126}.

\bibitem{Tsamis:1994ca}
N.~C. Tsamis and R.~P. Woodard, ``{Strong infrared effects in quantum
  gravity},''
\href{http://dx.doi.org/10.1006/aphy.1995.1015}{{\em Annals Phys.} {\bfseries
  238} (1995) 1--82}.

\bibitem{Tsamis:1996qm}
N.~C. Tsamis and R.~P. Woodard, ``{The Quantum gravitational back reaction on
  inflation},'' \href{http://dx.doi.org/10.1006/aphy.1997.5613}{{\em Annals
  Phys.} {\bfseries 253} (1997) 1--54},
\href{http://arxiv.org/abs/hep-ph/9602316}{{\ttfamily arXiv:hep-ph/9602316
  [hep-ph]}}.

\bibitem{Tsamis:1997za}
N.~C. Tsamis and R.~P. Woodard, ``{Matter contributions to the expansion rate
  of the universe},''
  \href{http://dx.doi.org/10.1016/S0370-2693(98)00159-2}{{\em Phys. Lett.}
  {\bfseries B426} (1998) 21--28},
\href{http://arxiv.org/abs/hep-ph/9710466}{{\ttfamily arXiv:hep-ph/9710466
  [hep-ph]}}.

\bibitem{Weinberg:2005vy}
S.~Weinberg, ``{Quantum contributions to cosmological correlations},''
  \href{http://dx.doi.org/10.1103/PhysRevD.72.043514}{{\em Phys. Rev.}
  {\bfseries D72} (2005) 043514},
\href{http://arxiv.org/abs/hep-th/0506236}{{\ttfamily arXiv:hep-th/0506236
  [hep-th]}}.

\bibitem{Weinberg:2006ac}
S.~Weinberg, ``{Quantum contributions to cosmological correlations. II. Can
  these corrections become large?},''
  \href{http://dx.doi.org/10.1103/PhysRevD.74.023508}{{\em Phys. Rev.}
  {\bfseries D74} (2006) 023508},
\href{http://arxiv.org/abs/hep-th/0605244}{{\ttfamily arXiv:hep-th/0605244
  [hep-th]}}.

\bibitem{Senatore:2009cf}
L.~Senatore and M.~Zaldarriaga, ``{On Loops in Inflation},''
  \href{http://dx.doi.org/10.1007/JHEP12(2010)008}{{\em JHEP} {\bfseries 12}
  (2010) 008},
\href{http://arxiv.org/abs/0912.2734}{{\ttfamily arXiv:0912.2734 [hep-th]}}.

\bibitem{Polyakov:2012uc}
A.~M. Polyakov, ``{Infrared instability of the de Sitter space},''
\href{http://arxiv.org/abs/1209.4135}{{\ttfamily arXiv:1209.4135 [hep-th]}}.

\bibitem{Senatore:2012ya}
L.~Senatore and M.~Zaldarriaga, ``{The constancy of $\zeta$ in single-clock
  Inflation at all loops},''
  \href{http://dx.doi.org/10.1007/JHEP09(2013)148}{{\em JHEP} {\bfseries 09}
  (2013) 148},
\href{http://arxiv.org/abs/1210.6048}{{\ttfamily arXiv:1210.6048 [hep-th]}}.

\bibitem{Assassi:2012et}
V.~Assassi, D.~Baumann, and D.~Green, ``{Symmetries and Loops in Inflation},''
  \href{http://dx.doi.org/10.1007/JHEP02(2013)151}{{\em JHEP} {\bfseries 02}
  (2013) 151},
\href{http://arxiv.org/abs/1210.7792}{{\ttfamily arXiv:1210.7792 [hep-th]}}.

\bibitem{Weinberg:2003sw}
S.~Weinberg, ``{Adiabatic modes in cosmology},''
  \href{http://dx.doi.org/10.1103/PhysRevD.67.123504}{{\em Phys. Rev.}
  {\bfseries D67} (2003) 123504},
\href{http://arxiv.org/abs/astro-ph/0302326}{{\ttfamily arXiv:astro-ph/0302326
  [astro-ph]}}.

\bibitem{Hinterbichler:2012nm}
K.~Hinterbichler, L.~Hui, and J.~Khoury, ``{Conformal Symmetries of Adiabatic
  Modes in Cosmology},''
  \href{http://dx.doi.org/10.1088/1475-7516/2012/08/017}{{\em JCAP} {\bfseries
  1208} (2012) 017},
\href{http://arxiv.org/abs/1203.6351}{{\ttfamily arXiv:1203.6351 [hep-th]}}.

\bibitem{Salopek:1990jq}
D.~S. Salopek and J.~R. Bond, ``{Nonlinear evolution of long wavelength metric
  fluctuations in inflationary models},''
\href{http://dx.doi.org/10.1103/PhysRevD.42.3936}{{\em Phys. Rev.} {\bfseries
  D42} (1990) 3936--3962}.

\bibitem{Creminelli:2004yq}
P.~Creminelli and M.~Zaldarriaga, ``{Single field consistency relation for the
  3-point function},''
  \href{http://dx.doi.org/10.1088/1475-7516/2004/10/006}{{\em JCAP} {\bfseries
  0410} (2004) 006},
\href{http://arxiv.org/abs/astro-ph/0407059}{{\ttfamily arXiv:astro-ph/0407059
  [astro-ph]}}.

\bibitem{Burgess:2010dd}
C.~P. Burgess, R.~Holman, L.~Leblond, and S.~Shandera, ``{Breakdown of
  Semiclassical Methods in de Sitter Space},''
  \href{http://dx.doi.org/10.1088/1475-7516/2010/10/017}{{\em JCAP} {\bfseries
  1010} (2010) 017},
\href{http://arxiv.org/abs/1005.3551}{{\ttfamily arXiv:1005.3551 [hep-th]}}.

\bibitem{Marolf:2010zp}
D.~Marolf and I.~A. Morrison, ``{The IR stability of de Sitter: Loop
  corrections to scalar propagators},''
  \href{http://dx.doi.org/10.1103/PhysRevD.82.105032}{{\em Phys. Rev.}
  {\bfseries D82} (2010) 105032},
\href{http://arxiv.org/abs/1006.0035}{{\ttfamily arXiv:1006.0035 [gr-qc]}}.

\bibitem{Marolf:2011sh}
D.~Marolf and I.~A. Morrison, ``{The IR stability of de Sitter QFT: Physical
  initial conditions},''
  \href{http://dx.doi.org/10.1007/s10714-011-1233-3}{{\em Gen. Rel. Grav.}
  {\bfseries 43} (2011) 3497--3530},
\href{http://arxiv.org/abs/1104.4343}{{\ttfamily arXiv:1104.4343 [gr-qc]}}.

\bibitem{Marolf:2012kh}
D.~Marolf, I.~A. Morrison, and M.~Srednicki, ``{Perturbative S-matrix for
  massive scalar fields in global de Sitter space},''
  \href{http://dx.doi.org/10.1088/0264-9381/30/15/155023}{{\em Class. Quant.
  Grav.} {\bfseries 30} (2013) 155023},
\href{http://arxiv.org/abs/1209.6039}{{\ttfamily arXiv:1209.6039 [hep-th]}}.

\bibitem{Anninos:2014lwa}
D.~Anninos, T.~Anous, D.~Z. Freedman, and G.~Konstantinidis, ``{Late-time
  Structure of the Bunch-Davies De Sitter Wavefunction},''
  \href{http://dx.doi.org/10.1088/1475-7516/2015/11/048}{{\em JCAP} {\bfseries
  1511} no.~11, (2015) 048},
\href{http://arxiv.org/abs/1406.5490}{{\ttfamily arXiv:1406.5490 [hep-th]}}.

\bibitem{Burgess:2015ajz}
C.~P. Burgess, R.~Holman, and G.~Tasinato, ``{Open EFTs, IR effects \&
  late-time resummations: systematic corrections in stochastic inflation},''
  \href{http://dx.doi.org/10.1007/JHEP01(2016)153}{{\em JHEP} {\bfseries 01}
  (2016) 153},
\href{http://arxiv.org/abs/1512.00169}{{\ttfamily arXiv:1512.00169 [gr-qc]}}.

\bibitem{Gorbenko:2019rza}
V.~Gorbenko and L.~Senatore, ``{$\lambda \phi^4$ in dS},''
\href{http://arxiv.org/abs/1911.00022}{{\ttfamily arXiv:1911.00022 [hep-th]}}.

\bibitem{Baumgart:2019clc}
M.~Baumgart and R.~Sundrum, ``{De Sitter Diagrammar and the Resummation of
  Time},''
\href{http://arxiv.org/abs/1912.09502}{{\ttfamily arXiv:1912.09502 [hep-th]}}.

\bibitem{Tsamis:2005hd}
N.~C. Tsamis and R.~P. Woodard, ``{Stochastic quantum gravitational
  inflation},'' \href{http://dx.doi.org/10.1016/j.nuclphysb.2005.06.031}{{\em
  Nucl. Phys.} {\bfseries B724} (2005) 295--328},
\href{http://arxiv.org/abs/gr-qc/0505115}{{\ttfamily arXiv:gr-qc/0505115
  [gr-qc]}}.

\bibitem{Riotto:2008mv}
A.~Riotto and M.~S. Sloth, ``{On Resumming Inflationary Perturbations beyond
  One-loop},'' \href{http://dx.doi.org/10.1088/1475-7516/2008/04/030}{{\em
  JCAP} {\bfseries 0804} (2008) 030},
\href{http://arxiv.org/abs/0801.1845}{{\ttfamily arXiv:0801.1845 [hep-ph]}}.

\bibitem{Seery:2009hs}
D.~Seery, ``{A parton picture of de Sitter space during slow-roll inflation},''
  \href{http://dx.doi.org/10.1088/1475-7516/2009/05/021}{{\em JCAP} {\bfseries
  0905} (2009) 021},
\href{http://arxiv.org/abs/0903.2788}{{\ttfamily arXiv:0903.2788
  [astro-ph.CO]}}.

\bibitem{Serreau:2013psa}
J.~Serreau and R.~Parentani, ``{Nonperturbative resummation of de Sitter
  infrared logarithms in the large-N limit},''
  \href{http://dx.doi.org/10.1103/PhysRevD.87.085012}{{\em Phys. Rev.}
  {\bfseries D87} (2013) 085012},
\href{http://arxiv.org/abs/1302.3262}{{\ttfamily arXiv:1302.3262 [hep-th]}}.

\bibitem{Nacir:2016fzi}
D.~López~Nacir, F.~D. Mazzitelli, and L.~G. Trombetta, ``{$O(N)$ model in
  Euclidean de Sitter space: beyond the leading infrared approximation},''
  \href{http://dx.doi.org/10.1007/JHEP09(2016)117}{{\em JHEP} {\bfseries 09}
  (2016) 117},
\href{http://arxiv.org/abs/1606.03481}{{\ttfamily arXiv:1606.03481 [hep-th]}}.

\bibitem{Tanaka:1975ti}
F.~Tanaka, ``{Coherent Representation of Dynamical Renormalization Group in
  Bose Systems},''
\href{http://dx.doi.org/10.1143/PTP.54.289}{{\em Prog. Theor. Phys.} {\bfseries
  54} (1975) 289--290}.

\bibitem{Boyanovsky:1998aa}
D.~Boyanovsky, H.~J. de~Vega, R.~Holman, and M.~Simionato, ``{Dynamical
  renormalization group resummation of finite temperature infrared
  divergences},'' \href{http://dx.doi.org/10.1103/PhysRevD.60.065003}{{\em
  Phys. Rev.} {\bfseries D60} (1999) 065003},
\href{http://arxiv.org/abs/hep-ph/9809346}{{\ttfamily arXiv:hep-ph/9809346
  [hep-ph]}}.

\bibitem{Boyanovsky:2003ui}
D.~Boyanovsky and H.~J. de~Vega, ``{Dynamical renormalization group approach to
  relaxation in quantum field theory},''
  \href{http://dx.doi.org/10.1016/S0003-4916(03)00115-5}{{\em Annals Phys.}
  {\bfseries 307} (2003) 335--371},
\href{http://arxiv.org/abs/hep-ph/0302055}{{\ttfamily arXiv:hep-ph/0302055
  [hep-ph]}}.

\bibitem{Boyanovsky:2004gq}
D.~Boyanovsky and H.~J. de~Vega, ``{Particle decay in inflationary
  cosmology},'' \href{http://dx.doi.org/10.1103/PhysRevD.70.063508}{{\em Phys.
  Rev.} {\bfseries D70} (2004) 063508},
\href{http://arxiv.org/abs/astro-ph/0406287}{{\ttfamily arXiv:astro-ph/0406287
  [astro-ph]}}.

\bibitem{McDonald:2006hf}
P.~McDonald, ``{Dark matter clustering: a simple renormalization group
  approach},'' \href{http://dx.doi.org/10.1103/PhysRevD.75.043514}{{\em Phys.
  Rev.} {\bfseries D75} (2007) 043514},
\href{http://arxiv.org/abs/astro-ph/0606028}{{\ttfamily arXiv:astro-ph/0606028
  [astro-ph]}}.

\bibitem{Podolsky:2008qq}
D.~I. Podolsky, ``{Dynamical renormalization group methods in theory of eternal
  inflation},'' \href{http://dx.doi.org/10.1134/S0202289309010174}{{\em Grav.
  Cosmol.} {\bfseries 15} (2009) 69--74},
\href{http://arxiv.org/abs/0809.2453}{{\ttfamily arXiv:0809.2453 [gr-qc]}}.

\bibitem{Burgess:2009bs}
C.~P. Burgess, L.~Leblond, R.~Holman, and S.~Shandera, ``{Super-Hubble de
  Sitter Fluctuations and the Dynamical RG},''
  \href{http://dx.doi.org/10.1088/1475-7516/2010/03/033}{{\em JCAP} {\bfseries
  1003} (2010) 033},
\href{http://arxiv.org/abs/0912.1608}{{\ttfamily arXiv:0912.1608 [hep-th]}}.

\bibitem{Maldacena:2011nz}
J.~M. Maldacena and G.~L. Pimentel, ``{On graviton non-Gaussianities during
  inflation},'' \href{http://dx.doi.org/10.1007/JHEP09(2011)045}{{\em JHEP}
  {\bfseries 09} (2011) 045},
\href{http://arxiv.org/abs/1104.2846}{{\ttfamily arXiv:1104.2846 [hep-th]}}.

\bibitem{Raju:2012zr}
S.~Raju, ``{New Recursion Relations and a Flat Space Limit for AdS/CFT
  Correlators},'' \href{http://dx.doi.org/10.1103/PhysRevD.85.126009}{{\em
  Phys. Rev.} {\bfseries D85} (2012) 126009},
\href{http://arxiv.org/abs/1201.6449}{{\ttfamily arXiv:1201.6449 [hep-th]}}.

\bibitem{Arkani-Hamed:2015bza}
N.~Arkani-Hamed and J.~Maldacena, ``{Cosmological Collider Physics},''
\href{http://arxiv.org/abs/1503.08043}{{\ttfamily arXiv:1503.08043 [hep-th]}}.

\bibitem{Arkani-Hamed:2018kmz}
N.~Arkani-Hamed, D.~Baumann, H.~Lee, and G.~L. Pimentel, ``{The Cosmological
  Bootstrap: Inflationary Correlators from Symmetries and Singularities},''
\href{http://arxiv.org/abs/1811.00024}{{\ttfamily arXiv:1811.00024 [hep-th]}}.

\bibitem{Arkani-Hamed:2018bjr}
N.~Arkani-Hamed and P.~Benincasa, ``{On the Emergence of Lorentz Invariance and
  Unitarity from the Scattering Facet of Cosmological Polytopes},''
\href{http://arxiv.org/abs/1811.01125}{{\ttfamily arXiv:1811.01125 [hep-th]}}.

\bibitem{Benincasa:2018ssx}
P.~Benincasa, ``{From the flat-space S-matrix to the Wavefunction of the
  Universe},''
\href{http://arxiv.org/abs/1811.02515}{{\ttfamily arXiv:1811.02515 [hep-th]}}.

\bibitem{Farnsworth:2017tbz}
K.~Farnsworth, M.~A. Luty, and V.~Prilepina, ``{Weyl versus Conformal
  Invariance in Quantum Field Theory},''
  \href{http://dx.doi.org/10.1007/JHEP10(2017)170}{{\em JHEP} {\bfseries 10}
  (2017) 170},
\href{http://arxiv.org/abs/1702.07079}{{\ttfamily arXiv:1702.07079 [hep-th]}}.

\bibitem{Polchinski:1983gv}
J.~Polchinski, ``{Renormalization and Effective Lagrangians},''
\href{http://dx.doi.org/10.1016/0550-3213(84)90287-6}{{\em Nucl. Phys.}
  {\bfseries B231} (1984) 269--295}.

\bibitem{Green:2013rd}
D.~Green, M.~Lewandowski, L.~Senatore, E.~Silverstein, and M.~Zaldarriaga,
  ``{Anomalous Dimensions and Non-Gaussianity},''
  \href{http://dx.doi.org/10.1007/JHEP10(2013)171}{{\em JHEP} {\bfseries 10}
  (2013) 171},
\href{http://arxiv.org/abs/1301.2630}{{\ttfamily arXiv:1301.2630 [hep-th]}}.

\bibitem{Behbahani:2012be}
S.~R. Behbahani and D.~Green, ``{Collective Symmetry Breaking and Resonant
  Non-Gaussianity},''
  \href{http://dx.doi.org/10.1088/1475-7516/2012/11/056}{{\em JCAP} {\bfseries
  1211} (2012) 056},
\href{http://arxiv.org/abs/1207.2779}{{\ttfamily arXiv:1207.2779 [hep-th]}}.

\end{thebibliography}\endgroup

\end{document}